\documentclass[longbibliography,
reprint,
superscriptaddress,
 amsmath,amssymb,
 aps,pre,
onecolumn
]{revtex4-2}
\usepackage{xcolor} 
\usepackage{setspace}
\usepackage{graphicx}% Include figure files
\usepackage{dcolumn}% Align table columns on decimal point
\usepackage{booktabs}
\usepackage[dvipsnames]{xcolor}
\usepackage{color}
\usepackage{tikz}
\usetikzlibrary{shapes}
\usepackage{soul}
\usepackage{hhline}

\usepackage{siunitx}

\usepackage{amsmath}
\usepackage{commath}
\usepackage{mathtools}
\usepackage[hidelinks, colorlinks=true, linkcolor=blue, citecolor=blue, filecolor=blue, urlcolor=blue]{hyperref}
\usepackage{bm}% bold math
\begin{document}
\setstretch{1.5}

\title{Rheology of dense suspensions of granular spherocylinders by particle-based simulation} 
\author{Alex Dixon}
\affiliation{School of Engineering, University of Edinburgh, Edinburgh EH9 3JL, United Kingdom}
\author{John Hone}
\affiliation{
Syngenta Jealott’s Hill International Research Centre,
Bracknell RG42 6EY,
United Kingdom
}
\author{Gavin Melaugh}
\affiliation{School of Engineering, University of Edinburgh, Edinburgh EH9 3JL, United Kingdom}
\affiliation{School of Physics and Astronomy, The University of Edinburgh, Edinburgh EH9 3FD, United Kingdom}
\author{Christopher Ness}%
\affiliation{School of Engineering, University of Edinburgh, Edinburgh EH9 3JL, United Kingdom}

\date{\today}

\begin{abstract}
Dense suspensions of rod-shaped granular particles are widespread in nature and manufacturing, where their fluid mechanical properties are often paramount. We have developed a particle-based simulation that models such suspensions under simple shear flow,  providing predictions of the viscosity and microstructure for a given solids volume fraction and particle aspect ratio. The model tracks the trajectories of spherocylindrical rods under the action of short-range frictional contact and hydrodynamic forces, inspired by similar tools that have generated new insight into suspensions of granular spheres. It incorporates new schemes for the computation of lubrication forces between spherocylinders and the dynamic determination of the timestep. For aspect ratios up to 20, the model predicts a viscosity spike at shear start-up, giving way to steady state viscosities that increase systematically with volume fraction and aspect ratio. Likewise, particle alignment increases with volume fraction up to an aspect-ratio-dependent critical point. Our model corroborates the limited experimental rheology data available for suspensions of granular rods, and offers a tool for fundamental exploration of the fluid mechanics, microstructure and rheology of this widespread material.
\end{abstract}

\maketitle
%\tableofcontents

\section{Introduction}

Suspensions of elongated, rod-shaped particles at high solid volume fraction appear in many natural and industrial settings,
where it is critical that their rheology and, ultimately, their fluid mechanical behaviour, is well characterised~\cite{butler2018microstructural}.
In forested areas,
for instance,
fallen trees and branches can accumulate on bodies of water with their structure evolving slowly and eventually forming sprawling interlocked `log-jam' structures at river narrowings,
negatively impacting animal habitats~\cite{deshpande2019logjams}. Elsewhere, magma flows often comprise crystalline needle-like rods of much smaller size but similar aspect ratio $A$ that generate complex flow phenomenology and lead to rapid arrest as the material cools~\cite{cimarelli2011rheology,moitra2015effects}. 
In industry,
widely employed reactive crystallisation processes critical to the manufacture of speciality chemicals generate concentrated suspensions of micron-sized shards or whiskers that can be well-approximated as rods~\cite{choo2019review, zhang2010synthesis, cao2022mechanical}.
Likewise, in paper making viscous slurries of fibres form entangled networks that give rise to non-Newtonian rheology~\cite{derakhshandeh2011rheology, jung2025entanglement}. 
In all of these cases, the system eventually reaches sufficiently high solid volume fraction that rod-rod interactions dominate the overall stress response
--
that is, they are \emph{dense}
--
and the constituent particles are large enough that diffusive motion originating from Brownian forces is significantly slower than the convective motion introduced by shearing. Additionally, the particles typically remain suspended at long times, either by neutral buoyancy with the fluid, by viscous resuspension, or by support through networks of contact forces. 

The basic understanding of suspension rheology has its origins in the work of Stokes~\cite{stokes1851},
Einstein~\cite{Einstein_1911} and Batchelor~\cite{batchelor1972determination}, who described how the presence of spheres enhances the viscosity under dilute and semi-dilute conditions.
The extension to dense conditions is usually empirical,
with the most common expression relating the suspension viscosity $\eta_\mathrm{s}$ to the volume fraction $\phi$
being the power law $\eta_\mathrm{s}\sim(1-\phi/\phi_m)^{-\varsigma}$ associated with~\citet{krieger1959mechanism}.
Here $\phi_m$ is the volume fraction at which the viscosity diverges, which is highly sensitive to microscopic details.
More recently,~\citet{boyer2011unifying} expanded upon this by introducing a model that unifies suspensions of non-Brownian, non-attractive particles with dry granular flows. They demonstrated that, in simple shear with rate $\dot{\gamma}$ and imposed particle pressure $P$, the macroscopic friction coefficient and the volume fraction depend only on the viscous number $J=\eta\dot{\gamma}/P$, where $\eta$ is the liquid phase viscosity. It follows that $\eta_\mathrm{s}$ depends only on $\phi$ and $\phi_m$, and is independent of $\dot{\gamma}$.
In principle this argument holds irrespective of particle shape. 

For non-spherical particles, the additional degree of freedom associated with their orientation introduces complexity.
In shear flows, an isolated slender rod will undergo Jeffery orbits~\cite{jeffery1922motion},
with period set by the shear rate, the particle aspect ratio $A$,
and the inclination out of the flow-gradient plane.
In sheared packings of dry rods, meanwhile, the time- and rod-averaged alignment is along the flow direction and makes a small angle with the streamlines~\cite{borzsonyi2012orientational}.
This angle decreases with increasing $A$,
meaning particles spend more time oriented along the flow direction.
Crucially,
this alignment lowers the effective instantaneous viscosity when compared to a randomly oriented assembly of the same rods at the same solids fraction~\cite{borzsonyi2012shear}.
The alignment of suspended particles can be markedly different, in some cases showing orientation along the vorticity direction during large amplitude oscillatory shear. This is thought to originate in a mechanism of collision minimisation, facilitated by confinement introduced at the walls~\cite{franceschini2011transverse, snook2012vorticity, franceschini2014dynamics}.
Thus,
particle shape leads to interesting microstructural phenomenology with the result that,
unlike for the idealised case of non-Brownian spheres,
knowledge of $\phi$ alone is insufficient for rheological prediction.

In general, though, suspensions of rods exhibit a steady increase in $\eta_\mathrm{s}$ with $\phi$, as with spheres, and this effect is enhanced as $A$ grows~\cite{ralambotiana1997viscosity, bibbo1987rheology, mackaplow1996numerical, chaouche2001rheology, tapia2017rheology,snook2014normal}.
Most experiments have focused on the dilute or semi-dilute regimes, up to around $\phi=0.2$, especially for large-$A$ rods, with only limited experimental results beyond this point due to the difficulty of preparing samples and measuring the rheology in such systems (but see \emph{e.g.}~\cite{tapia2017rheology}).
As a result, the rheological properties of the dense regime, where contacts and near-contact hydrodynamics dominate, remains incompletely characterised despite its importance to the application areas mentioned above.
Simulations offer a complementary approach that permits controlled exploration of $\phi$, $A$, orientational order and other microscopic details (for instance friction or adhesion) that are difficult to access experimentally.

There have been many attempts to simulate suspensions of rods,
spanning continuum~\cite{strand1987computation, quinones2025smoluchowski},
particle-based~\cite{mackaplow1996numerical, yamane1994numerical}
and
coupled particle-fluid~\cite{khan2023rheology} approaches.
%The focus of this paper is on particle-based simulations.
Stokesian dynamics (SD)~\cite{brady1988stokesian} is a particle-based model that includes long-range many-body hydrodynamics and short-range lubrication and has successfully been used to model dilute-to-semi dilute suspensions in the limit of zero inertia~\cite{phung1996stokesian, foss2000structure, singh2000normal, kutteh2004methods}.
Using a simplified SD approach,
\citet{mari2020shear} simulated suspensions of short rods, treating them as dimers formed from two overlapping spheres.
Near jamming, frictionless rods align along the vorticity (rather than the flow) direction.
Going beyond the overlapping spheres approach, \citet{yamane1994numerical} first derived a model for the lubrication forces between rod-shaped particles,
limited to normal interactions between non-parallel rod shafts (omitting parallel shafts as well as shaft-end and end-end pairs).
This was later expanded by~\citet{butler2002dynamic}
for spherocylinders, to include interactions with the hemispherical ends of the rods.
\citet{fan1998direct} included the
non-parallel shaft-shaft term~\cite{yamane1994numerical} when simulating the motion of rigid, non-Brownian particles modelled as cylinders in a sheared suspension,
including both short and long-range interactions.
Doing so they reproduced the dependence of $\eta_\mathrm{s}$ on $\phi$ and $A$ within the dilute to semi-dilute regime, in agreement with experimental data (although at a high computational cost).
More broadly, for large $\phi$ where interparticle interactions dominate, fully resolving the many contacts and lubrication interactions between rods with SD can become prohibitively expensive~\cite{durlofsky1987dynamic} due to computationally expensive matrix inversions required to obtain particle velocities. 

An alternative particle-based technique to study the dense regime is the discrete element method (DEM), originally developed for dry granular materials~\cite{cundall1979discrete}.
Here the zero inertia constraint is relaxed,
and particle trajectories are updated directly following Newton's second law, allowing the use of much cheaper velocity-Verlet timestepping algorithms.
This approach has been used widely for modelling both spherical and non-spherical particles~\cite{plimpton1995lammps, yousefian2022orientational, zhang2025hindered, trulsson2021directional, bilotto2025shear},
and its variants are becoming a mainstay of granular suspension modelling~\cite{ness2023simulating,dong2020unifying,ge2020implementation,gallier2014rheology,more2021unifying}.
There is an extensive literature on such models applied to dry granular rods,
which are most commonly described as chains of spheres~\cite{guo2012numerical,pol2023unified,nan2015simulation}. Following this approach,
rods are predicted to align with their principal axes at a small angle to the flow direction,
leading to shear-induced nematic order consistent with experimental observation~\cite{guo2015computational, borzsonyi2012orientational, borzsonyi2012shear}.
Similar models have also reproduced frictional fibres jamming at high volume fraction, forming a dense contact network~\cite{guo2015computational}. 
Other DEM studies have modelled rods directly as ellipsoids~\cite{yousefian2022orientational,bilotto2025shear}, cylinders~\cite{berzi2016stresses, bilotto2025shear} or spherocylinders~\cite{nath2019rheology} and together begin to reveal how particle shape and friction affect the alignment of particles and bulk rheology.
Collectively,
these works show that velocity fluctuations follow the kinetic theory prediction up to a critical packing, demonstrating that granular gas models can be applicable beyond spheres~\cite{berzi2022dense}.
They further show that frictional rods shear thicken at $\phi = 0.54-0.59$,
just below the jamming point at $\phi_m$,
with
larger $A$ associated with a stronger viscosity increase, and frictionless systems showing only shear thinning~\cite{nath2019rheology}.
Long ($A = 4$) and flat ($A = 0.25$) cylinders were found to align with the shear direction (with frictional systems having weaker induced order due to frictional torques generating relative rotations between particles), whereas the directors of the near-spherical cylinders ($A = 0.8-1.0$) remained disordered~\cite{berzi2022dense,berzi2016stresses,nath2019rheology}.
More recently, these models have been applied to suspensions,
with~\citet{anzivino2024shear} simulating systems of frictionless rod-sphere mixtures in suspension,
finding that adding short rods reduces the mixture viscosity, whereas adding long rods increases it. 
These and other particle-based simulation studies highlight that anisotropic particles can greatly modify the rheology compared to spheres, but they also reveal limitations. Most models for elongated particles in the literature either treat rods as chains of spheres or idealise their interactions, often without the necessary hydrodynamic forces needed to simulate suspensions. 

In this paper we present a particle-based simulation for dense suspensions of non-bendy, non-Brownian spherocylinders
that borrows elements from the above works to produce a model that balances a faithful rendering of the dominant microscopic physics with computational tractability.
Alongside frictional particle contact forces,
fluid drag and shear-induced lift,
we adapt the lubrication forces of~\citet{yamane1994numerical} and \citet{butler2002dynamic} by developing a scheme that avoids discontinuities when traversing particle-pair configurations.
We track the rod trajectories using a conventional DEM algorithm complemented by a novel dynamic timestep calculation, operating in a regime where particle inertia is present but negligible.
For suspensions of spheres,
models containing these microscopic physics~\cite{cheal2018rheology,seto2018normal,trulsson2012transition} have proven instrumental in corroborating and extending the rheological model proposed by \citet{boyer2011unifying}.
Our model described in the following seeks to significantly extend this understanding to the broad class of suspensions of rod-shaped particles.
Using the model, we advance the fundamental understanding of the rheology of suspensions of granular rods
by demonstrating
(i) the dependence of the viscosity $\eta_\mathrm{s}$ on the volume fraction $\phi$ (spanning $\phi\approx0$ to $\phi_m$) and particle aspect ratio (spanning $A=1$ to $20$),
and 
(ii) the complex relation between $\phi$, $A$, and the particle alignment, measured using the orientational order parameter.
This sets the scene for future studies that map, for instance, the full dependence of $\phi_m$ on the microstructural details, the relation between $\phi$, $A$ and the normal stresses, and the behaviour under inhomogeneous conditions, all of which are crucial for a full fluid mechanical description of dense granular rod suspensions.

\section{\label{sec:level2}Simulation Method}

\subsection{Equations of Motion of Suspended Granular Rods}

\begin{figure}
        \centering
        \includegraphics[trim = 4.3cm 54.9cm 54.7cm 6.9cm, clip,width=0.8\textwidth]{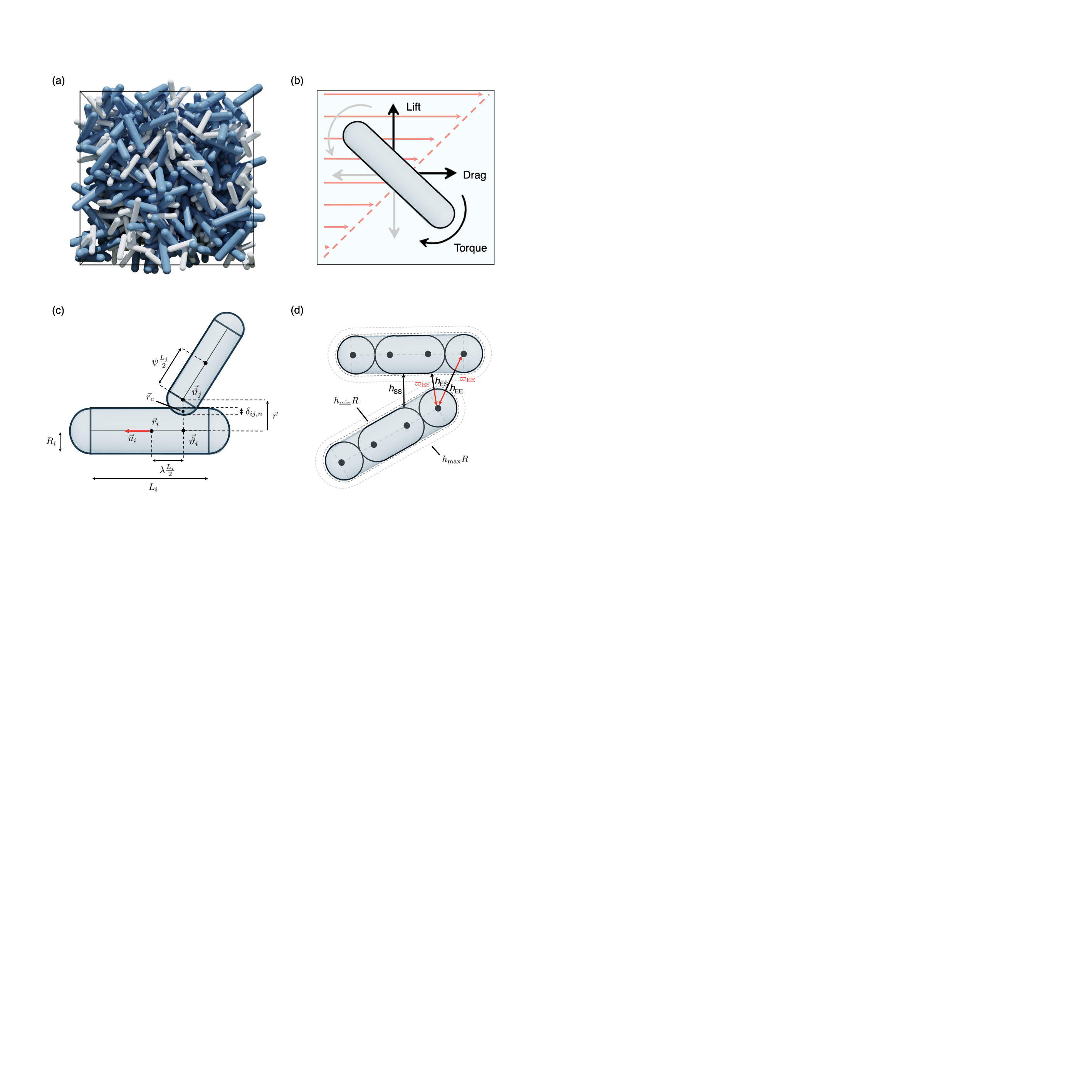}
        \caption{
        Particle-based simulation of suspensions of sheared granular spherocylindrical rods. 
        Shown in (a) is a render of a typical simulation for aspect ratio $A = 5$, solid volume fraction $\phi = 0.3$, with $N=1000$ particles coloured according to their radii with $R$ in white and $1.4R$ in blue.
        The terms present in the equation of motion of the particles are
        (b) single-body hydrodynamics comprising drag, lift and torque, governed by the imposed linear velocity profile sketched in red;
        (c) repulsive contact forces between overlapping pairs, labelled here to illustrate terms present in the interaction detection algorithm used to find pairwise contacts;
        (d) lubrication forces between neighbouring particles, calculated using a blending approach to incorporate end-end, end-shaft and shaft-shaft interactions.
        The various values labelled in (c) and (d) are defined in the main text.
        }
        \label{fig_1}
\end{figure}

We model particles as cylinders with hemispherical caps, Figure~\ref{fig_1}, hereafter referred to as spherocylinders. This geometry resembles rod-shaped particles but simplifies interaction detection compared to alternatives such as cylinders~\cite{gan2020simulation} or spheroids~\cite{lin2002distance}, although it is limited to prolate particles. Each particle $i$ is defined by its centre of mass $\Vec{r}_i$, radius $R_i$ (the same for the shaft and hemispherical caps), shaft length $L_i$ and direction unit vector $\Vec{u}_i$.
We define the aspect ratio as $A=({L+2R})/{2R}$ (so spheres have $A=1$).
In what follows, an arrow above a variable denotes that it is a 3D vector.
The translational motion of particle $i$ is described by the following equation
\begin{equation}
\label{trans motion}
    m_i \frac{d \Vec{v}_i}{dt} =  \sum_j\Vec{F}_{\mathrm{C}, ij} + \sum_k\Vec{F}_{\mathrm{L}, ik} + \Vec{F}_{\mathrm{H}, i},
\end{equation}
where $m_i$ is the mass of the particle,
$\Vec{v}_i$ is its translational velocity,
and $t$ is time.
The forces included in the model are pairwise contact forces $\Vec{F}_{\mathrm{C}, ij}$ acting between particle $i$ and its contacting neighbours $j$,
pairwise lubrication forces $\Vec{F}_{\mathrm{L}, ik}$ acting between particle $i$ and its near-neighbours $k$,
and single-body hydrodynamic forces $\Vec{F}_{\mathrm{H}, i}$.
The rotational motion of each particle is similarly governed by
\begin{equation}
\label{ang motion}
    \mathbf{I_i} \frac{d \Vec{\omega}_i}{dt} =  \sum_j\Vec{T}_{\mathrm{C}, ij} + \sum_k\Vec{T}_{\mathrm{L}, ik} + \Vec{T}_{\mathrm{H}, i},
\end{equation}
where $\mathbf{I_i}$ is a diagonal matrix denoting the mass moment of inertia of particle $i$ (throughout, bold variables represent $3\times3$ tensors), $\Vec{\omega}_i$ is its angular velocity, and $\Vec{T}$ the torque acting on it, which similarly has contributions from contacts, lubrication and single-body hydrodynamics labelled equivalently to the forces.
For a pairwise interaction between particles $i$ and $j$,
the torque on particle $i$ is calculated as
$\vec{T_i} = \vec{F}_{ij} \times \vec{\ell}_{ij}$,
where $\vec{\ell}_{ij}$ is the moment arm from the centre of mass $\vec{r}_i$ to the point of interaction along rod $i$ and $\Vec{F}_{ij}$ is the relevant force term.
After summing the forces and torques on each particle, details of which are described in turn below, the equations of motion are integrated numerically using a standard implementation of the Velocity-Verlet algorithm.

\subsection{Single-Body Hydrodynamic Forces}

\label{subsec-Hydroforces}

Suspended neutrally buoyant particles experience hydrodynamic forces when their velocity differs from that of their surrounding fluid. 
Due to their anisotropic shape, suspended rods experience orientation-dependent driving,
so that the forces $\vec{F}_\mathrm{H}$ have contributions from both drag $\vec{F}_\mathrm{H,D}$ and lift $\vec{F}_\mathrm{H,L}$ and there also exists a torque contribution $\vec{T}_\mathrm{H}$.
These are illustrated in Figure~\ref{fig_1}(b).
Although numerical simulations have sought to model the hydrodynamic forces on spherocylinders for limited aspect ratios and finite Reynolds number~\cite{cheron2024drag, feng2023general},
a generalised analytical solution does not yet exist for the drag forces experienced in Stokes flow, the regime of interest here.
In lieu of this, we adopt the result for prolate spheroids, which exists in the literature in analytical form and represents a reasonable approximation for spherocylindrical rods,
especially at higher aspect ratios~\cite{happel2012low}. The following summarises the work of~\citet{zhang2001ellipsoidal}. Here single-hatted variables are defined in the particle frame of reference, which has origin at the centre of mass of the particle $\vec{r}_i$ and $z$-axis aligned with the principal axis of the particle $\vec{u}_i$. Unhatted variables are defined in the lab frame of reference.
% A double hat is used for variables defined in the co-moving frame of reference, which has axes parallel to the lab frame and its origin placed at $\vec{r}_i$.
The drag force experienced by axisymmetric particles was derived by~\citet{brenner1963stokes} and takes a similar form to the Stokes drag on a sphere~\cite{stokes1851}
\begin{equation}
    \Vec{F}_{\mathrm{H,D},i} = \mathbf{K} \pi \eta R_i \cdot (\Vec{v}_{\mathrm{f},i} - \Vec{v}_i)\mathrm{.}
\end{equation}
Here $\eta$ is the liquid phase viscosity, $\vec{v}_{\mathrm{f},i}$ is the fluid velocity at the centre of mass of particle $i$ assuming the velocity profile to be linear across the domain, and $\mathbf{{K}}$ is a geometric resistance tensor that takes into account the orientation of the particle with respect to the background fluid velocity field.
The geometric resistance tensor is first defined as a diagonal tensor in the particle frame of reference 
\cite{fan1995sublayer} as
% %
\begin{subequations}
\begin{align}
    \hat{k}_{\hat{x}\hat{x}} &= \hat{k}_{\hat{y}\hat{y}} = \frac{16(A^2 - 1)}{\left[ (2A^2 - 3) \frac{\ln{\left( A + \sqrt{A^2 - 1}\right)}}{\sqrt{A^2 - 1}} \right] + A}, \\[10pt]
    \hat{k}_{\hat{z}\hat{z}} &= \frac{8(A^2 - 1)}{\left[ (2A^2 - 1) \frac{\ln{\left( A + \sqrt{A^2 - 1}\right)}}{\sqrt{A^2 - 1}} \right] - A}.
\end{align}
\end{subequations}
Here $\hat{k}_{\hat{z}\hat{z}}$ corresponds to the principal axis along the length of a rod,
while the degenerate terms $\hat{k}_{\hat{x}\hat{x}}$ and $\hat{k}_{\hat{y}\hat{y}}$ represent the short axes. This can then be transformed to the lab frame of reference ($\mathbf{{K}}  = \mathbf{{M}}^{-1} \mathbf{\hat{{K}}} \mathbf{{M}}$) using the transformation matrix $\mathbf{M}$
\begin{equation}
\mathbf{M} = 
\begin{bmatrix}
\cos\xi \cos\kappa - \cos\theta \sin\kappa \sin\xi & 
\cos\xi \sin\kappa + \cos\theta \cos\kappa \sin\xi & 
\sin\xi \sin\theta \\
-\sin\xi \cos\kappa - \cos\theta \sin\kappa \cos\xi & 
-\sin\xi \sin\kappa + \cos\theta \cos\kappa \cos\xi & 
\cos\xi \sin\theta \\
\sin\theta \sin\kappa & 
-\sin\theta \cos\kappa & 
\cos\theta
\end{bmatrix},
\end{equation}
where $\xi$, $\kappa$ and $\theta$ are the Euler angles between the particle director $\vec{u}_i$ and the lab frame.
For axisymmetric particles degeneracy about the principal axis allows us to set $\xi=0$ so that $\mathbf{M}$ simplifies to
\begin{equation}
\mathbf{M} = 
\begin{bmatrix}
\cos\kappa & 
\sin\kappa & 
0 \\
- \cos\theta \sin\kappa & 
\cos\theta \cos\kappa & 
\sin\theta \\
\sin\theta \sin\kappa & 
-\sin\theta \cos\kappa & 
\cos\theta
\end{bmatrix}.
\end{equation}
In addition to the drag force, a particle may experience shear-induced lift $\vec{F}_{\mathrm{H,L}}$ in the presence of a velocity profile.
Lift is an inertial effect
so is included in our model for completeness.
As will be explained in Section~\ref{simsetup},
however,
our simulation parameters are set such that particle inertia is negligible so that in practice lift forces are significantly smaller than all other forces in the system.
The lift acting on an arbitrary shaped particle \cite{harper1968maximum} with a fluid velocity gradient ${d{v}_x}/{dz}$ in the $z$-direction is
\begin{equation}
    \vec{F}_{\mathrm{H,L},i} = \pi^2 R_i^2\sqrt{{\eta\rho_{\mathrm{F}}}}
    \frac{d{v}_x / d z}{\sqrt{\abs{d {v}_x / d z}}}(\mathbf{{K}} \cdot \mathbf{L}_{\mathrm{M}} \cdot \mathbf{{K}}) \cdot (\vec{v}_{\mathrm{f},i} - \vec{v}_i),
\end{equation}
where $\rho_{\mathrm{F}}$ is the fluid density and $\mathbf{L}_\mathrm{M}$ is the coefficient matrix of the lift tensor for arbitrary shapes~\cite{harper1968maximum} and widely used for spheroids and spherocylinders~\cite{wang2018lift, cheron2024drag, zhang2001ellipsoidal, cui2018constitutive}
\begin{equation}
    \mathbf{L}_\mathrm{M} = 
    \begin{bmatrix}
        0.0501 & 0.0329 & 0.00 \\
        0.0182 & 0.0173 & 0.00 \\
        0.00 & 0.00 & 0.0373
    \end{bmatrix}.
\end{equation}
The hydrodynamic torque on an ellipsoidal particle in a linear shear flow is given by~\citet{jeffery1922motion}, wherein the total torque is composed of a viscous drag component that resists rotation and an additional shear-induced component arising from the imposed velocity gradient:
\begin{subequations}
\begin{align}
    \vec{T}_{\mathrm{H}, x} &= \frac{16 \pi \eta R^3 A}{3 \left(\beta_0 + A^2 \gamma_0 \right)} \left[ (1 - A^2) d_{\hat{z}\hat{y}} + (1 + A^2)(w_{\hat{z}\hat{y}} - \omega_{\hat{x}}) \right], \\[10pt]
    \vec{T}_{\mathrm{H}, y} &= \frac{16 \pi \eta R^3 A}{3 \left(\alpha_0 + A^2 \gamma_0 \right)} \left[ (A^2 - 1) d_{\hat{x}\hat{z}} + (1 + A^2)(w_{\hat{x}\hat{z}} - \omega_{\hat{y}}) \right], \\[10pt]
    \vec{T}_{\mathrm{H}, z} &= \frac{32 \pi \eta R^3 A}{3 \left(\alpha_0 + \beta_0 \right)} (w_{\hat{y}\hat{x}} - \omega_{\hat{z}}),
\end{align}
\end{subequations}
where $\omega$ as before is the angular velocity of the particle (with $i$ subscripts dropped for clarity). Additionally, $d$ are elements of the deformation rate tensor \(\mathbf{\hat{D}} = \frac{1}{2} \left( \nabla v + (\nabla v)^T \right)\) and $w$ are elements of the spin tensor \(\mathbf{\hat{W}} = \frac{1}{2} \left( \nabla v - (\nabla v)^T \right)\), both written in the particle frame of reference. The velocity gradient tensor, naturally expressed in the lab frame, can be transformed into the particle frame via $\hat{\mathbf{G}}=\mathbf{M}\mathbf{G}\mathbf{M}^{-1}$.

The dimensionless parameters $\alpha_0$, $\beta_0$ and $\gamma_0$~\cite{gallily1979orderly} take into account the geometry of the particles and are given by
\begin{subequations}
\begin{align}
    \alpha_0 = \beta_0 &= \frac{A^2}{A^2 - 1} + \frac{A}{2 (A^2 - 1)^{3/2}} \ln{\left[ \frac{A - \sqrt{A^2 - 1}}{A + \sqrt{A^2 - 1}} \right]}, \\[10pt]
    \gamma_0 &= -\frac{2}{A^2 - 1} - \frac{A}{(A^2 - 1)^{3/2}} \ln{\left[ \frac{A - \sqrt{A^2 - 1}}{A + \sqrt{A^2 - 1}} \right]}.
\end{align}
\end{subequations}
This shear-induced torque captures the characteristic rotational behaviour of elongated particles in viscous flows, reproducing the Jeffery orbits observed in real suspensions~\cite{jeffery1922motion}.

\subsection{Pairwise Interaction Forces}
\subsubsection{Interaction Detection}
Before introducing the contact and lubrication forces, we first describe our algorithm for detecting whether neighbouring particles are close enough to be interacting.
Spherocylinders have both a long and short dimension, meaning the minimum distance between their surfaces cannot be determined from the positions of their centres of mass. Instead, detecting a contact requires computing the shortest distance between the two rod \emph{spines} and comparing it to the sum of their radii.
To do this,
we implement the method described by \citet{vega1994fast} for rods of arbitrary radii ($R_i$, $R_j$) although alternative methods have been used~\cite{pournin2005three, angklomkleaw2019simulation}.
In general,
points along the spines of rods $i$ and $j$ can be defined as
% $\vec{S}_i$ and $\vec{S}_i$ corresponding to the closest point between the two particles is defined as,
%
\begin{subequations}
\label{min_dist}
\begin{align}
\Vec{\vartheta}_i &=  \Vec{r}_i + \lambda \frac{L_i}{2} \Vec{u}_i,
\label{eq:min_dist_a} \\[10pt]
\Vec{\vartheta}_j &=  \Vec{r}_j + \psi \frac{L_j}{2} \Vec{u}_j,\label{eq:min_dist_b}
\end{align}
\end{subequations}
where $\lambda$ and $\psi$ are fractions of the half length of a rod that signify the distances along the spine from the centres of mass $\Vec{r}_i$ and $\Vec{r}_j$ respectively.
These distances are sketched in Figure~\ref{fig_1}(c).
The values of $\psi$ and $\lambda$ are $[-1, 1]$ for finite length rods.
The vector between these points $\vec{r}$ is then
\begin{equation}
    \Vec{r} = \Vec{\vartheta}_j - \Vec{\vartheta}_i = \Vec{r}_{ij} + \psi \frac{L_j}{2} \Vec{u}_j - \lambda \frac{L_i}{2} \Vec{u}_i,
\end{equation}
where $\vec{r}_{ij} = \vec{r}_j - \vec{r}_i$ is the vector pointing between the two centres of mass.
The values of $\lambda$ and $\psi$ that minimise $\abs{\vec{r}}$ (denoted $\lambda'$, $\psi'$) also minimise $\vec{r}^2$,
and for infinitely long rods these can be found by taking the derivative of $\vec{r}^2$ with respect to $\lambda$ and $\psi$,
which leads to
\begin{subequations}
\begin{align}
\lambda' &= \frac{2}{L_i} \frac{(\Vec{r}_{ij} \cdot \Vec{u}_i) - (\Vec{u}_i \cdot \Vec{u}_j)(\Vec{r}_{ij} \cdot \Vec{u}_j)}{1 - (\Vec{u}_i \cdot \Vec{u}_j)^2},  \\[10pt]
\psi' &= - \frac{2}{L_j} \frac{(\Vec{r}_{ij} \cdot \Vec{u}_j) - (\Vec{u}_i \cdot \Vec{u}_j)(\Vec{r}_{ij} \cdot \Vec{u}_i)}{1 - (\Vec{u}_i \cdot \Vec{u}_j)^2}.
\end{align}
\end{subequations}
These minimising values can be $[-\infty, \infty]$. If both are $[-1, 1]$ then the closest point between the infinitely extended rod spines falls within the finite bounds of the actual rods and $\lambda'$ and $\psi'$ can be used directly in Equation~\ref{min_dist} to find the points of closest approach.
If one of the fractions does not fall within $[-1, 1]$, then an additional step is required to find the closest distance within the finite bounds of the rod.
First, whichever of the two fractions is further from the limits $[-1, 1]$ is trimmed to be at the limit (\emph{e.g.} $\lambda' = -10.2$ becomes $\lambda'' = -1$). This truncated value ($\lambda''$ or $\psi''$) is then used to find the value of the complementary fraction ($\psi''$ or $\lambda''$),
again from the derivative of ${\vec{r}}^2$,
using the appropriate one of the following expressions
\begin{subequations}
\begin{align}
\lambda'' &= \psi'' \frac{L_j}{2} (\Vec{u}_i \cdot \Vec{u}_j) + (\Vec{r}_{ij} \cdot \Vec{u}_i), \\[10pt]
\psi'' &= \lambda'' \frac{L_i}{2} (\Vec{u}_i \cdot \Vec{u}_j) - (\Vec{r}_{ij} \cdot \Vec{u}_j).
\end{align}
\end{subequations}
If the value thus obtained is within the bounds $[-1, 1]$,
then $\lambda''$ and $\psi''$ can directly replace $\lambda$ and $\psi$ in Equation~\ref{min_dist}.
Otherwise,
the obtained value is itself trimmed to within the bounds,
and can then be used.

Once $\vec{\vartheta}_i$ and $\vec{\vartheta}_j$ are found,
a determination can be made of whether two particles are interacting.
Particles are in direct contact if the distance between the two closest points along their spines,
$\vec{r}$,
is less than the sum of their radii,
that is, $\abs{\vec{r}} \leq R_i + R_j$.
This is true for spherocylinders since
the vector connecting the closest points on the two rods is always normal both to the spines and to the particle surfaces.
The overlap between two particles is then $\delta_{n, ij} = \left( R_i + R_j \right) - \abs{\vec{r}}$,
where the normal direction of the contact vector is given by $\vec{n}_{ij} = \vec{r}/\abs{\vec{r}}$.
The point of interaction $\vec{r}_c$ is then taken as the midpoint between the surfaces of the interacting particles. 

\subsubsection{Contact Forces}
Direct contacts between particles are modelled as Hookean with Coulombic friction, with normal and tangential components of the forces computed as follows, based on the work by \citet{mahajan2018non}. 
The total contact force acting on particle $i$ is the sum of the normal $\Vec{F}_{\mathrm{C}, n, ij}$ and tangential $\Vec{F}_{\mathrm{C}, t,ij}$ components contributed by contacting neighbours $j$
\begin{equation}
    \Vec{F}_{\mathrm{C}, ij} = \sum (\Vec{F}_{\mathrm{C}, n, ij} + \Vec{F}_{\mathrm{C}, t, ij}).
\end{equation}
Here the normal force is modelled as a damped spring:
\begin{equation} \label{norma_fo}
    \Vec{F}_{\mathrm{C},n, ij} = -k_n \delta_{n, ij} \Vec{n}_{ij} + \zeta_n \Vec{v}_{n, ij},
\end{equation}
where $k_n$ is the normal spring constant, $\vec{n}_{ij}$ is the unit direction vector of the interaction as defined in the preceding section, $\zeta_n$ is the normal damping coefficient, and recalling that $\delta_{n, ij}$ is the scalar overlap between the contacting particles.
Here $\vec{v}_{n, ij}$ is the component of the relative velocity at the point of contact along the direction $\vec{n}_{ij}$, defined as
\begin{equation}
\label{relveleq}
    \Vec{v}_{ij} = \Vec{v}_j - \Vec{v}_i + \Vec{\omega}_j \times (\Vec{r}_{c} - \Vec{r}_j) - \Vec{\omega}_i \times (\Vec{r}_c - \Vec{r}_i).
\end{equation}
The normal damping coefficient~\cite{pournin2005behavior} entering Equation~\ref{norma_fo} is calculated as
\begin{equation}
     \zeta_n = - \frac{2 m_\mathrm{eff}}{t_c} \ln{e_n},
\end{equation}
where $m_\mathrm{eff}$ is the effective mass of the particle pair participating in the interaction ($m_\mathrm{eff} = \frac{m_i m_j}{m_i+m_j}$) and $t_c$ is the duration of the contact determined as
\begin{equation}
\label{collsion time}
    t_c = \sqrt{\frac{m_\mathrm{eff}}{k_n} (\pi^2 + \ln{e_n}^2)}.
\end{equation}
Here $e_n$ is the normal restitution coefficient of the particle,
which sets how much kinetic energy is preserved after an interaction in the absence of suspending fluid.

The tangential contact force is modelled similarly,
but with the inclusion of an upper limit set by a friction coefficient $\mu$ and defined according to
\begin{equation}
    \Vec{F}_{\mathrm{C}, t, ij} = \begin{cases}
    -k_t \Vec{\delta}_{t, ij} + \zeta_t \Vec{v}_{t, ij} & \mathrm{for } \abs{\Vec{F}_{\mathrm{C},t, ij}} \leq \mu \abs{\Vec{F}_{\mathrm{C}, n, ij}}, \\[5pt]
    - \mu \abs{\Vec{F}_{\mathrm{C}, n, ij}} \Vec{t}_{ij} & \mathrm{for } \abs{\Vec{F}_{\mathrm{C},t, ij}} > \mu \abs{\Vec{F}_{\mathrm{C}, n,ij}},
    \end{cases}
\end{equation}
where $\Vec{\delta}_{t, ij}$ is the tangential overlap which quantifies the elastic tangential deformation of a particle since the onset of the contact and $k_t$ is the tangential spring constant. Here $\zeta_t$ is the tangential damping coefficient, $\vec{v}_{t, ij}$ is the tangential component of the relative velocity at the point of contact and $\vec{t}_{ij}$ is the unit vector in the direction of the static tangential force, ensuring the sliding force opposes the tangential deformation and relative motion.
The quantities $t_c$, $e_n$ and $e_t$ can in principle be measured from simple single contact experiments,
informing an appropriate choice for $k_n$.
Having specified the tangential restitution coefficient $e_t$ and calculated the contact time $t_c$, the tangential stiffness $k_t$ can then be found using~\cite{pournin2005behavior}
\begin{equation}
    k_t = t_c^{-2}  \left( \frac{1}{m_\mathrm{eff}} + \frac{\abs{\Vec{r}_c - \Vec{r}_i}^2}{\langle I_i \rangle} + \frac{\abs{\Vec{r}_c - \Vec{r}_j}^2}{\langle I_j \rangle} \right)^{-1} (\pi^2 + \ln{e_t}^2),
\end{equation}
where $\langle I_i \rangle$ is the average mass moment of inertia defined below in Equation \ref{inertiaeqts} and $e_t$ is the tangential restitution.
The tangential damping coefficient $\zeta_t$~\cite{pournin2005behavior} is
\begin{equation}
    \zeta_t = t_c^{-1}  \left( \frac{1}{m_\mathrm{eff}} + \frac{\abs{\Vec{r}_c - \Vec{r}_i}^2}{\langle I_i \rangle} + \frac{\abs{\Vec{r}_c - \Vec{r}_j}^2}{\langle I_j \rangle} \right)^{-1}  \ln{e_t}.
\end{equation}
For a spherocylinder the non-zero components of the mass moments of inertia $\mathbf{\hat{I}}$, in the particle frame of reference, can be calculated~\cite{pournin2005three,constantin2015aspect} as 
\begin{subequations}
\label{inertiaeqts}
\begin{align}
    I_{\hat{x}\hat{x}} &= I_{\hat{y}\hat{y}} = \pi \rho_{\mathrm{P}} \left(\frac{1}{12} R^2 L^3 + \frac{8}{15} R^5 + \frac{1}{3} R^3 L^2+ \frac{3}{4} R^4 L\right), \\[10pt]
    I_{\hat{z}\hat{z}} &= \pi \rho_{\mathrm{P}} \left(\frac{1}{2} R^4 L + \frac{8}{15} R^5\right), \\[10pt]
    \langle I \rangle &= \frac{1}{3} (I_{\hat{x}\hat{x}} + I_{\hat{y}\hat{y}} + I_{\hat{z}\hat{z}}),
\end{align}
\end{subequations}
where $\rho_{\mathrm{P}}$ is the particle density.
The tangential overlap $\Vec{\delta}_{t, ij}$ is obtained by integrating the relative tangential velocity between two particles over the duration of the contact
\begin{equation}
    \Vec{\delta}_{t, ij}(t) = \int_{t_{c, 0}}^{t} \Vec{v}_{t, ij} dt.
\end{equation}
Together, the normal and tangential force models described above provide a description of particle contact interactions within the simulation, capturing both damped spring forces and frictional dissipation.

\subsubsection{Lubrication forces}
\label{lubforce_section}
Lubrication forces arise from short-range hydrodynamic interactions generated when two particles within a fluid move relative to each other in close proximity.
These forces originate in the viscosity of the fluid in the narrow gap between the particles,
opposing their relative motion.
Lubrication forces become particularly important at small surface-to-surface separations $h$ and hence are an essential component of the equation of motion for particles in dense suspension.
Importantly,
the discontinuity in the curvature of spherocylinders at the end-shaft junction precludes an analytical model for the lubrication force~\cite{janoschek2013accurate}.
To address this,
~\citet{butler2002dynamic} proposed a modification of the \citet{yamane1994numerical} model,
adapting it to spherocylinders by treating the three possible interaction configurations separately, namely the shaft-shaft (SS), end-shaft (ES) and end-end (EE) interactions
\begin{subequations}
\label{lub_forces}
\begin{align}
\vec{F}_{\mathrm{L,SS},ij} &= -24 \frac{\eta \pi R^2}{h_{ij}} \frac{A^2}{(2A + 1) \left[ \left( A^2 + \tfrac{1}{4} \right) \abs{\vec{u}_i \times \vec{u}_j}^2 + A \left( 1 + (\vec{u}_i \cdot \vec{u}_j)^2 \right) \right]^{1/2}} \vec{v}_{n, ij} \vec{n}_{ij}, 
\label{lub_forces:ss}
\\[10pt]
\vec{F}_{\mathrm{L,ES},ij} &= - \frac{4}{\sqrt{2}} \frac{\eta \pi R^2}{h_{ij}} \vec{v}_{n, ij} \vec{n}_{ij}, 
\label{lub_forces:se}
\\[10pt]
\vec{F}_{\mathrm{L,EE},ij} &= - \frac{3}{2} \frac{\eta \pi R^2}{h_{ij}} \vec{v}_{n, ij} \vec{n}_{ij},
\label{lub_forces:ee}
\end{align}
\end{subequations}
where $A$ is the aspect ratio, $h$ is the separation between the surfaces of particles ($h_{ij} = \abs{\vec{r}} - \left( R_i + R_j \right)$) and $\eta$ is the fluid viscosity.
This separate treatment of the different interaction modes leads to a discontinuity in the force when a pair of particles moves in time from one type of interaction to another,
an issue that is amplified at higher aspect ratios.
We resolve this by computing all three components (Equations~\ref{lub_forces:ss} - \ref{lub_forces:ee}) for all neighbouring pairs,
but assigning weights to each term dependent on the relative position of the two particles.
This requires three lengths to be defined that quantify the SS, ES and EE distances, with the importance of each force term then being inversely proportional to its associated distance.
The initial shaft-shaft weight is thus defined as 
$
C_{\mathrm{SS}} = (1/\varpi_\mathrm{SS})/(1/\varpi_\mathrm{SS}+1/\varpi_\mathrm{ES}+1/\varpi_\mathrm{EE})\mathrm{,}
$
with analogous expressions for $C_\mathrm{ES}$ and $C_\mathrm{EE}$
so that by construction we have
$C_{\mathrm{SS}} + C_{\mathrm{ES}} + C_{\mathrm{EE}} = 1$.
These weights contribute to the overall expression for the lubrication force as
\begin{equation}
    \vec{F}_{\mathrm{L}, ij} = C_{\mathrm{SS}} \vec{F}_{\mathrm{L, SS}, ij} + C_{\mathrm{ES}} \vec{F}_{\mathrm{L, ES}, ij} + C_{\mathrm{EE}} \vec{F}_{\mathrm{L, EE}, ij}.
    \label{lub_forces:total}
\end{equation}
The distances $\varpi_\mathrm{SS}$ etc. must be defined such that the appropriate component dominates, \emph{i.e.} $C_{\mathrm{EE}}$ is most influential in an end-end interaction and $C_{\mathrm{SS}}$ is most influential when rods are parallel.
This is achieved by imagining the rod as split into spherical end sections with a smaller internal cylindrical shaft section. Within this representation we define the distances $\varpi_\mathrm{EE}=h_\mathrm{EE}+2R$, $\varpi_\mathrm{ES}=h_\mathrm{ES}+R$ and $\varpi_\mathrm{SS}=h_\mathrm{SS}$ as sketched in Figure~\ref{fig_1}(d).
Here each $h$ term is used for the respective force calculation in Equation~\ref{lub_forces},
while the corresponding $\varpi$ term is used to set the weighting assigned to that force.
This scheme alone resolves the aforementioned discontinuities, but to further improve the correction the weights can be manually adjusted to produce smoother transitions between interaction types (\emph{e.g.} $aC_{\mathrm{SS}} + bC_{\mathrm{ES}} + cC_{\mathrm{EE}} = 1$, where $a$, $b$ and $c$ are manually set corrective values and within our simulations are $a = 0.1$, $b = 1.0$, $c = 5.0$).
These values are determined by considering a series of different primitive pairwise motions (\emph{i.e.} particles sliding past one another along a common axis; perpendicular to one another and so on) that involve transitions in dominance from one type of lubrication interaction to another (\emph{e.g.} transitioning from shaft-shaft to shaft-end).
These weight correctors are then individually adjusted until a smooth profile is produced for each case.
Having obtained these profiles,
we then verify that the choice of weights has minimal impact on the rheological predictions of our model.
This approach is successful for aspect ratios $A\geq3$,
since the middle section needs to be at minimum a sphere for the distances to be calculated.
When evaluating each of the force components in Equation~\ref{lub_forces:total}, the corresponding separation $h_\mathrm{SS}$ etc. is used along with the average $R$ for interactions between particles of different sizes.

To ensure numerical stability as the gap between particle surfaces vanishes,
a minimum separation is specified.
This distance is defined as a fixed fraction $h_\mathrm{min}$ of the radii for an interacting pair, so that below separations of $h_{\mathrm{min}}(R_1+R_2)$ the separation $h_{ij}$ used in calculations isn't further decreased.
Such a lengthscale might physically represent the size of asperities on the particle surfaces, which are not explicitly modelled here.
Additionally, to reduce computational cost and avoid unphysical interactions with secondary neighbours (\emph{e.g.} where particles interact through other particles) a maximum separation $h_{\mathrm{max}}$ was similarly introduced to define a point above which lubrication forces between particles are not computed.
These lengths are sketched in Figure~\ref{fig_1}(d).

\subsection{Additional Model Details}

\subsubsection{Stress Calculation}

The stress within the system is calculated as the sum of the stress contributions from pairwise particle interactions $\boldsymbol{\sigma}_\mathrm{P}$, and a fluid contribution $\boldsymbol{\sigma}_\mathrm{F}$. The former is a $3\times3$ tensor constructed from the moment arm $\vec{\ell}$ and force $\vec{F}$ vectors, and is defined as~\cite{marschall2019shear}
\begin{equation}
    \boldsymbol{\sigma}_\mathrm{P} = -\frac{1}{V} \left( \sum^{N_\mathrm{C}}\vec{\ell}_{ij} \otimes \vec{F}_{\mathrm{C}, ij} + \sum^{N_\mathrm{L}} \vec{\ell}_{ik} \otimes \vec{F}_{\mathrm{L}, ik} ,\right)
\end{equation}
where $N_\mathrm{C}$ is the number of 
contacts (which occur between particles $i$ and $j$),
$N_\mathrm{L}$ is the number of lubrication interactions (which occur between particles $i$ and $k$),
$V$ is the total volume of the system,
and $\otimes$ is the outer product.
In the following we consider only the shear stress,
written for simplicity as $\sigma_\mathrm{P}$.
Meanwhile
the single-body forces generate an additional contribution to the shear stress
(labelled here as $\sigma_\mathrm{F}$ to denote the fluid contribution), given for spherical particles by~\citet{Einstein_1911} and later modified for spherocylinders by~\citet{kuhn1933quantitative} to read 
\begin{equation}
\label{fluid_visc}
{\sigma}_\mathrm{F} = \eta\dot{\gamma} \left( 1 + \frac{5}{2} \phi + \frac{\phi}{16} A^2\right)\mathrm{,}
\end{equation}
where we omit higher order terms in $\phi$.
This equation is strictly valid only in dilute systems,
and is limited in that it does not account for particle ordering,
instead assuming an isotropic and randomly distributed suspension.
At the high volume fractions of interest,
however, the fluid contribution becomes negligible relative to the stresses arising from particle-particle interactions.
Any contribution to the fluid stress coming from particle structure, following \emph{e.g.}~\citet{dinh1984rheological},
would not change this,
and would likely be dominated by structural contributions to the stress arising from lubrication and contact forces.
In the following we report the reduced shear viscosity of the suspension
as $\eta_\mathrm{s}=(\sigma_\mathrm{F}+\sigma_\mathrm{P})/\eta\dot{\gamma} =  \sigma/\eta\dot{\gamma}$.

\subsubsection{Quantifying Alignment}

To understand the material response to flow,
we characterise the microstructural arrangement of the particles
borrowing insight from the Q-tensor commonly used to characterise alignment in nematic liquid crystals.
We quantify the degree of collective ordering of the particles by computing the largest eigenvalue $S$ of the order tensor $\mathbf{T}$, the components of which are
defined as
\begin{equation}
\label{oop}
{T_{\alpha\beta}} = \left\langle \frac{3}{2} u_{i, \alpha} u_{i, \beta} - \frac{1}{2} \delta_{\alpha \beta}\right\rangle.
\end{equation}
Here $\alpha$ and $\beta$ represent the Cartesian coordinates, and angle brackets indicate averaging over all particles $i=1,\dots,N$.
This leads to the orientational order parameter $S$, a scalar metric that takes values in the range [0,1].
A value of $S=1$ means that the system is perfectly regularly aligned, \emph{i.e.} all particles are parallel to one another,
whereas a value of $S=0$ represents completely uncorrelated particle directors.
The eigenvector $\vec{u}_\mathrm{princ}$ associated with $S$ represents the principal direction of alignment.

\subsubsection{Simulation Setup and Parameters}
\label{simsetup}

 Particles governed by the above set of forces and torques are placed
 with initially random positions and orientations in a cubic domain consisting of periodic boundaries in the $x$ and $y$ directions and Lees-Edwards boundary conditions~\cite{lees1972computer} across the $z$ axis.
 An example snapshot of the simulated system is given in Figure~\ref{fig_1}(a).
The latter is used to introduce a velocity difference across the domain that leads, through the imposed drag forces described above, to a linear fluid velocity profile $v_{\mathrm{f},x}=\dot{\gamma}z$.
The minimum domain size is then $2\times$ the maximum interaction length of a particle.
Here this is equal to the total length of the particle plus the additional length of influence due to the lubrication force $h_\mathrm{max}R$ on each end.
The minimum system size is therefore $2R \left( A + 1 + h_\mathrm{max} \right)$.
It is standard practice in spheres to include a bidisperse mixture of radii (typically 1:1.4) to prevent crystallisation \cite{VanderWerf_Jin_Shattuck_OHern_2018, Ohern2002-tw, Zhang2009-yx,singh2024rheology}. 
For each value of $A$ we therefore simulated a bidisperse system comprising equal numbers of particles with radius and length ratio both set to 1:1.4. 
It is worth noting though that this choice has little impact on the rheology reported in the following.
Indeed we found little change in the steady state viscosity predicted by the simulation when moving from mono- to bi- to tridisperse systems. 

The parameters within the model are chosen such that (i) particle inertia can be considered insignificant $\dot\gamma^* \equiv \rho_{\mathrm{F}} \Dot{\gamma} R^2 / \eta < 10^{-2}$;
and (ii) rods can be assumed to be very hard $\dot{\gamma} \sqrt{\rho_{\mathrm{P}} R^3 / k_n} < 10^{-4}$.
Table~\ref{tab:glob-params} includes the material properties and other parameter values used throughout all of the simulations.
We chose modest values of the friction coefficient and normal restitution coefficient, noting that experimentally a wide range of values both above~\cite{mair2002influence} and below~\cite{foerster1994measurements} our chosen friction coefficient $\mu$ are observed in similar systems. The same is true for the normal restitution coefficient $e_n$~\cite{foerster1994measurements, joseph2001particle}.
We opted for values comfortably within the ranges found in the literature, acknowledging that they may not be representative of a specific single material.
Simulations suggest that the qualitative alignment behaviour reported here is independent of the particle friction, though we defer a systematic study of the role of the friction coefficient to future work.
Table~\ref{tab:sim-params} then gives the values that are altered for each test case. For each set of parameters presented the simulation was repeated for five realisations,
each starting from a distinct initial configuration.

\begin{table}
    \centering
    \caption{Global parameters used within all simulations reported in Figures~\ref{fig_2}-\ref{fig_6}.}
    \label{tab:glob-params}
    \setlength{\tabcolsep}{10pt}
    \begin{tabular}{
        c  % radius1
        c  % radius2
        c  % radiusratio
        c  % density f
        c  % spring_const
        c  % norm rest
        c  % tang rest
        c  % fric coef
        c  % viscosity
        c  % h min
        c  % h max
    }
    \toprule
    {$R_1$} & {$R_2$} & {$R_1:R_2$} & {$\rho_{\mathrm{F}}$}, {$\rho_{\mathrm{P}}$} & {$k_n$} & {$e_n$} & {$e_t$} & {$\mu$} & {$\eta$} & {$h_{\min}$} & {$h_{\max}$} \\ 
    {[L]} & {[L]} & [-] & {[ML$^{-3}$]} & {[MT$^{-2}$]} & {[ - ]} & {[ - ]} & {[ - ]} & {[ML$^{-1}$T$^{-1}$]} & {[ - ]}  & {[ - ]} \\
    \midrule
    % Example row (you can delete it)
    1 & 1.4 & 1:1 & 1 & $1.5\times10^6$ & 0.5 & 0.5 & 0.5 & 15 & 0.025 & 0.9 \\
    \bottomrule
    \end{tabular}
\end{table}

\begin{table}
  \centering
  \caption{Simulation parameters for each test case reported in Figures~\ref{fig_2}-\ref{fig_6}.}
  \label{tab:sim-params}
  \setlength{\tabcolsep}{10pt}
  \begin{tabular}{
    c % number of particles
    c   % vol frac
    c    % aspect ratio
    c   % shear rate
    c    % shear stress
  }
  \toprule
  $N$ & $\phi$ & {$A$} & {$\dot\gamma$} & {$\sigma_{\mathrm{fixed}}$} \\
  {[ - ]} & {[ - ]} & {[ - ]} & {[T$^{-1}$]} & {[ML$^{-1}$T$^{-2}$]}\\

  \midrule

    1000 & 0.01 & 5 & 0.07 & -  \\
    1000 & 0.1 & 5 & 0.07 & -  \\
    1000 & 0.2 & 5 & 0.07 & - \\
    1000 & 0.3 & 5 & 0.07 & - \\
    1000 & 0.4 & 5 & 0.07 & - \\
    1000 & 0.47 & 5 & 0.07 & - \\
    1000 & 0.5 & 5 & 0.07 & - \\
    1000 & 0.52 & 5 & 0.07 & - \\
    1000 & 0.55 & 5 & 0.07 & - \\
    
    1000 & 0.3 & 5 & 0.035 & - \\
    1000 & 0.3 & 5 & - & 2.5 \\
    1000 & 0.3 & 5 & - & 5 \\

    1000 & 0.01 & 10 & 0.07 & - \\
    1000 & 0.15 & 10 & 0.07 & - \\
    1000 & 0.25 & 10 & 0.07 & - \\
    1000 & 0.33 & 10 & 0.07 & - \\
    1000 & 0.4 & 10 & 0.07 & - \\
    1000 & 0.45 & 10 & 0.07 & - \\
    1000 & 0.5 & 10 & 0.07 & - \\
    1059 & 0.53 & 10 & 0.07 & - \\
    1119 & 0.56 & 10 & 0.07 & - \\

    1000 & 0.01 & 15 & 0.07 & - \\
    1000 & 0.13 & 15 & 0.07 & - \\
    1000 & 0.2 & 15 & 0.07 & - \\
    1144 & 0.28 & 15 & 0.07 & - \\
    1349 & 0.33 & 15 & 0.07 & - \\
    1553 & 0.38 & 15 & 0.07 & - \\
    1839 & 0.45 & 15 & 0.07 & - \\
    2003 & 0.49 & 15 & 0.07 & - \\
    2085 & 0.51 & 15 & 0.07 & - \\

    1000 & 0.01 & 20 & 0.07 & - \\
    1000 & 0.13 & 20 & 0.07 & - \\
    1660 & 0.24 & 20 & 0.07 & - \\
    1937 & 0.28 & 20 & 0.07 & - \\
    2284 & 0.33 & 20 & 0.07 & - \\
    2423 & 0.35 & 20 & 0.07 & - \\
    2631 & 0.38 & 20 & 0.07 & - \\

  \bottomrule
  \end{tabular}
\end{table}

\subsubsection{Time step size}

The equations of motion given above are integrated in time to update the particle trajectories.
The discrete time step $dt$ used for this integration needs to be carefully chosen to resolve all interactions while not adding unnecessarily to the computation time.
To achieve this we therefore need to approximate the characteristic timescale of each interaction.
The timescale associated with the contact force is the contact time $t_c$ (Equation~\ref{collsion time}). This timescale depends only on the material properties and the mass of the particles.
It increases with aspect ratio
but does not depend on the relative orientation of the contacting particles so is constant throughout an interaction.
There is also a timescale associated with the lubrication force $t_{\mathrm{lub}, \vec{F}}$. This can be approximated from the term in the equation of translational motion (Equation~\ref{trans motion}) corresponding to the lubrication force.
For simple shear we can make the assumption that the magnitude of the relative velocity between interacting particles is $\abs{dv} \sim R\dot\gamma$. For simplicity we may also assume that $\vec{F}_{\mathrm{L},\mathrm{SS}}$ is the dominant component of the lubrication force, since this grows with $A$.
With these assumptions we can approximate the timescale of a lubrication interaction as
\begin{equation}
\label{lubFtime_real}
    t_{\mathrm{lub}, \vec{F}, ij} \approx \frac{1}{36} \frac{\rho_{\mathrm{P}} h_{\mathrm{SS}} R^2 \dot\gamma}{\abs{\vec{v}_{n, ij}}\eta} \frac{(2A+1)(3A-1)}{A^2} \left[ \left( A^2 + \tfrac{1}{4} \right) \abs{\vec{u}_i \times \vec{u}_j}^2 + A \left( 1 + (\vec{u}_i \cdot \vec{u}_j)^2 \right) \right]^{1/2}.
\end{equation}
Here, unlike the contact time, we have a timescale that is configuration-dependent and therefore changes over the course of an interaction.
Moreover,
unlike $t_c$ the lubrication timescale decreases with increasing aspect ratio.
For elongated particles, the timescale for the force differs notably from that of the torque due to the large moment arm. The timescale associated with the lubrication torque $t_{\mathrm{lub}, \vec{T}}$ can be approximated from the equation of rotational motion for the lubrication torque (Equation~\ref{ang motion}).
For simple shear we can make the assumption that $\abs{d\omega} \approx \dot\gamma$. As $\vec{T} = \vec{F} \times \vec{\ell}$ and assuming $\abs{\vec{\ell}} \approx 0.5 L$, we can use the approximation $\abs{\vec{T}_{\mathrm{lub}}} \approx 0.5 L \abs{\vec{F}_{\mathrm{lub}, \mathrm{SS}}}$. The smallest component of $\mathbf{\hat{I}}$ is $I_{\hat{z}\hat{z}}$ for any aspect ratio and will lead to the limiting (smallest) timescale. With these assumptions we can approximate the timescale of the lubrication torque as
\begin{equation}
\label{lubTtime_real}
    t_{\mathrm{lub}, \vec{T}, ij} \approx \frac{1}{360} \frac{\rho_{\mathrm{P}} h_{\mathrm{SS}} R^2 \dot\gamma}{\abs{\vec{v}_{n,ij}} \eta } \frac{(2A+1)(15A-7)}{A^2(A-1)} \left[ \left( A^2 + \tfrac{1}{4} \right) \abs{\vec{u}_i \times \vec{u}_j}^2 + A \left( 1 + (\vec{u}_i \cdot \vec{u}_j)^2 \right) \right]^{1/2}.
\end{equation}
Again we have a timescale that changes over the course of an interaction and that now scales approximately with $1/A^2$.
The value of $dt$ for the subsequent time step is then chosen as the smallest of these timescales,
which we determined dynamically by 
inserting the actual values of $h_{\mathrm{SS}}$, $\vec{v}_{n,ij}$, $\vec{u}_i$ and $\vec{u}_j$ for each interaction into 
Equations~\ref{collsion time}, \ref{lubFtime_real} and \ref{lubTtime_real}.
 This ensures that $dt$ is always small enough to resolve ongoing interactions but never smaller than it needs to be,
 allowing for faster simulations.

 \begin{figure}
        \centering
        \includegraphics[trim = 1.6cm 77.7cm 37.4cm 2.7cm, clip,width=0.98\textwidth]{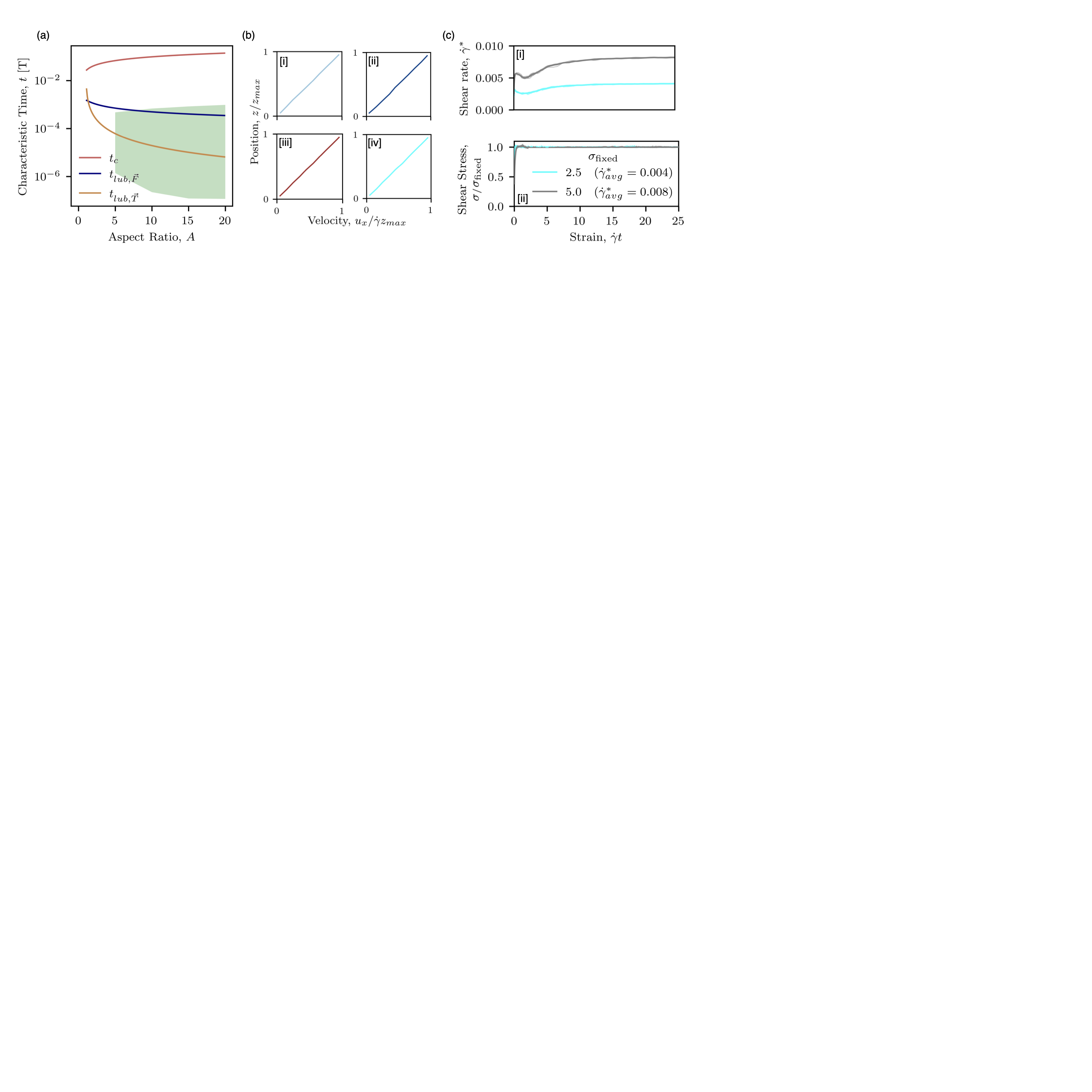}
        \caption{
        Results of prerequisite calculations and diagnostic simulations used to determine parameter values.
        (a) Shown by solid lines are the characteristic timescales associated with contact and lubrication interactions, defined in Equations~\ref{collsion time},~\ref{lub_force_time} and~\ref{lub_torque_time},
        while the area shaded in green highlights the range of dynamically adjusted time step sizes $dt$ that were used during the simulations; 
        (b) Example plots of the rescaled steady-state velocity profiles, demonstrating that the desired linear profiles, specified through the imposed shear rate $\dot{\gamma}$, are achieved.
        Shown are
        [i] $A=5$,
        $\phi=0.1$,
        $\dot\gamma^*=0.007$;
        [ii] $A=5$,
        $\phi=0.5$,
        $\dot\gamma^*=0.007$;
        [iii] $A=5$,
        $\phi=0.3$,
        $\dot\gamma^*=0.0035$;
        [iv] $A=5$,
        $\phi=0.3$,
        $\sigma_\mathrm{fixed} = 2.5$ [ML$^{-1}$T$^{-2}$] ($\dot\gamma^*_{\mathrm{avg}} = 0.004$);
        (c) Demonstration of the shear stress control algorithm described in the text,
        showing transients of:
        [i] the
        shear rate $\dot{\gamma}^*$ and
        [ii] the shear stress $\sigma$ rescaled by its set point values, for $\sigma_\mathrm{fixed}=\{2.5,5\}$ [ML$^{-1}$T$^{-2}$] (leading to $\dot\gamma^*_{\mathrm{avg}}$ = \{0.004, 0.008\}).
        }
        \label{fig_2}
\end{figure}

Alternatively we can calculate a static value of $dt$ by estimating the largest force or torque present when hydrodynamics dominate.
For the lubrication interactions this occurs when two parallel rods interact at their minimum distance.
In this configuration we can assume that
$h_{\mathrm{SS}} = Rh_{\mathrm{min}}$,
$\vec{u}_i \times \vec{u}_j = 0$
and
$\vec{u}_i \cdot \vec{u}_j = 1$.
For simple shear we can additionally assume
$\vec{v}_{n, ij} = AR \dot{\gamma}$.
(This latter assumption will likely break down at higher volume fractions,
where relative velocity is governed by contact forces (set by $k_n$) rather than the fluid flow (set by $\eta$),
leading to a smaller timescale than that calculated.)
Substituting these assumptions into Equations~\ref{lubFtime_real} and~\ref{lubTtime_real} we get
\begin{equation}
\label{lub_force_time}
    t_{\mathrm{lub}, \vec{F}} \approx \frac{\sqrt{2}}{36} \frac{\rho_{\mathrm{P}} h_{\mathrm{min}} R^2}{\eta} \frac{(2A+1)(3A-1)}{A^{5/2}},
\end{equation}
\begin{equation}
\label{lub_torque_time}
    t_{\mathrm{lub}, \vec{T}} \approx \frac{\sqrt{2}}{360} \frac{\rho_{\mathrm{P}} h_{\mathrm{min}} R^2}{\eta} \frac{(2A+1)(15A-7)}{A^{5/2}(A-1)}.
\end{equation}
These timescales are plotted as functions of $A$ in Figure~\ref{fig_2}(a) along with a shaded area denoting the range of actual timescales (obtained \emph{via} Equations~\ref{lubFtime_real} and~\ref{lubTtime_real}) used in each simulation.

\subsubsection{System Control Modes}

As described in Section~\ref{subsec-Hydroforces}, the fluid velocity profile $\vec{v}_\mathrm{f}(z)$ determines the single-body hydrodynamic forces on each particle. Throughout we prescribe a linear velocity profile by specifying a spatially invariant value of the gradient as $\dot{\gamma}=\partial v_\mathrm{f,x}/\partial z$.
The simulation can impose this shear rate under two modes.
In the first, we simply prescribe a fixed value of $\dot{\gamma}$ and run the dynamics.
In the second, we seek to model an imposed shear stress rather than an imposed shear rate.
To do so, at each time step $t$ we calculate the overall shear stress in the system as $\sigma_t$, and compare it to a specified set point shear stress $\sigma_\mathrm{fixed}$.
The offset is then used to adjust the imposed shear rate at the next time step $\dot{\gamma}_{t+\Delta t}$ (introducing a gain parameter $K_p$) according to
\begin{equation}
    \dot{\gamma}_{t+\Delta t} = \dot{\gamma}_{t} + \frac{\sigma_{\mathrm{fixed}} - \sigma_{t}}{\sigma_{\mathrm{fixed}}} K_p\mathrm{.}
\end{equation}
Setting {$K_p\approx10^{-3}\dot{\gamma}$}, and having verified that this does not introduce any unwanted transients or periodicity in the results,
this control scheme keeps the overall shear stress close to the set point value.

Example velocity profiles generated by
binning particle velocities along $z$ are plotted in Figure~\ref{fig_2}(b),
demonstrating that
the shear rate is kept close to the one prescribed,
irrespective of the volume fraction, shear rate or simulation mode (shear rate vs shear stress control). 
The performance of the stress controller for fixed shear stresses = \{2.5, 5\}[ML$^{-1}$T$^{-2}$]($\dot\gamma^*_{\mathrm{avg}}$ = \{0.004, 0.008\}) is shown in Figure~\ref{fig_2}(c).
The set point stress is closely maintained, showing an average error of $< 1 \%$ once reaching the set point.

\section{Results and Discussion}

\begin{figure}
        \centering
        \includegraphics[trim = 2.0cm 47.6cm 51cm 2cm, clip,width=0.7\textwidth]{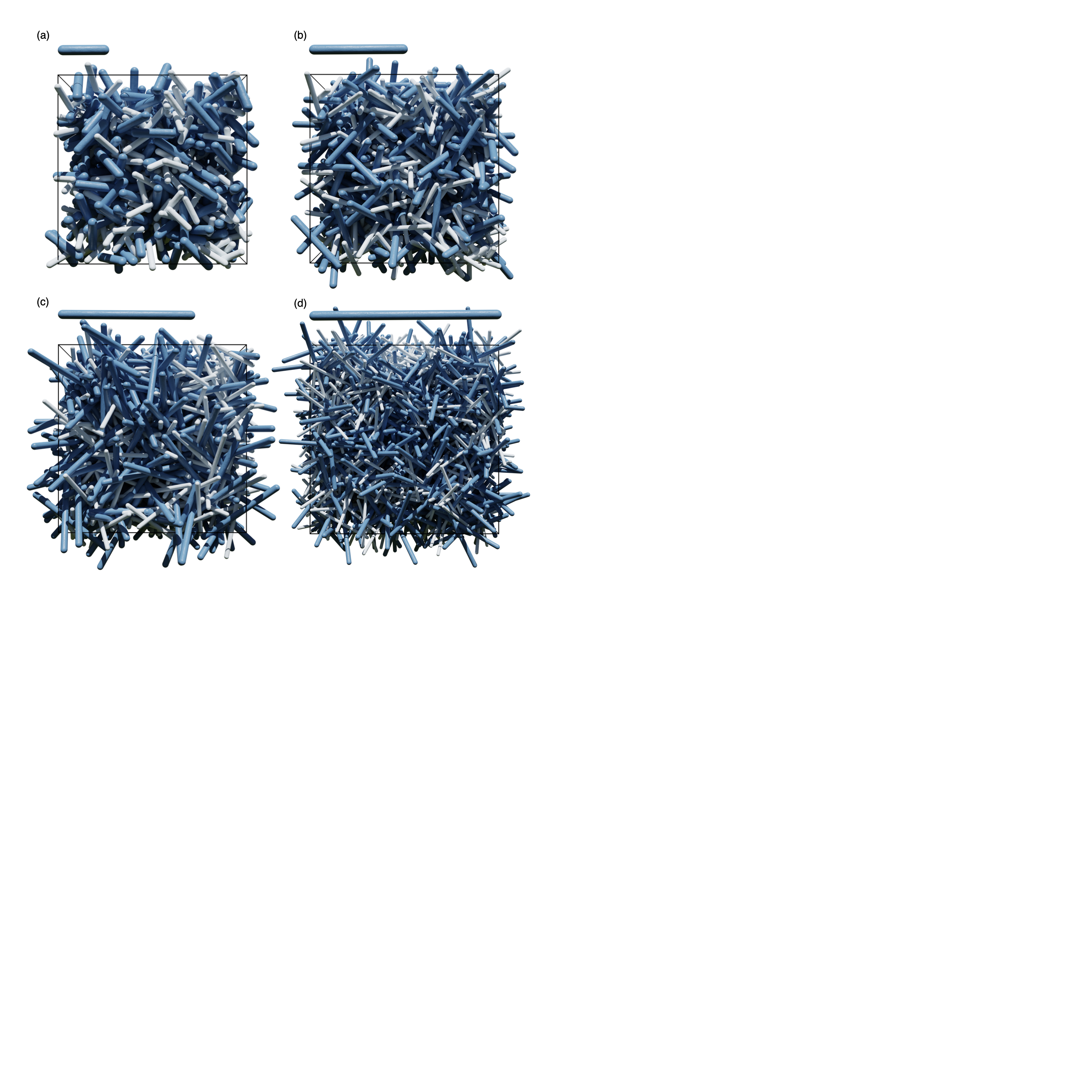}
        \caption{
        Simulation snapshots of the initial configurations of granular rod suspensions prior to shearing.
        Shown are systems at $\dot{\gamma}t=0$, with
        (a) $A=5, \phi =0.3$;
        (b) $A=10, \phi = 0.33$;
        (c) $A=15, \phi = 0.33$;
        (d) $A=20, \phi = 0.33$;
        where the particles are coloured based on radius with $R=1$ particles coloured white and $R=1.4$ particles coloured blue.
        }
        \label{fig_3}
\end{figure}

\subsection{Effect of volume fraction and aspect ratio on suspension viscosity}

\begin{figure}
        \centering
        \includegraphics[trim = 1.0cm 53.6cm 56cm 2cm, clip,width=0.8\textwidth]{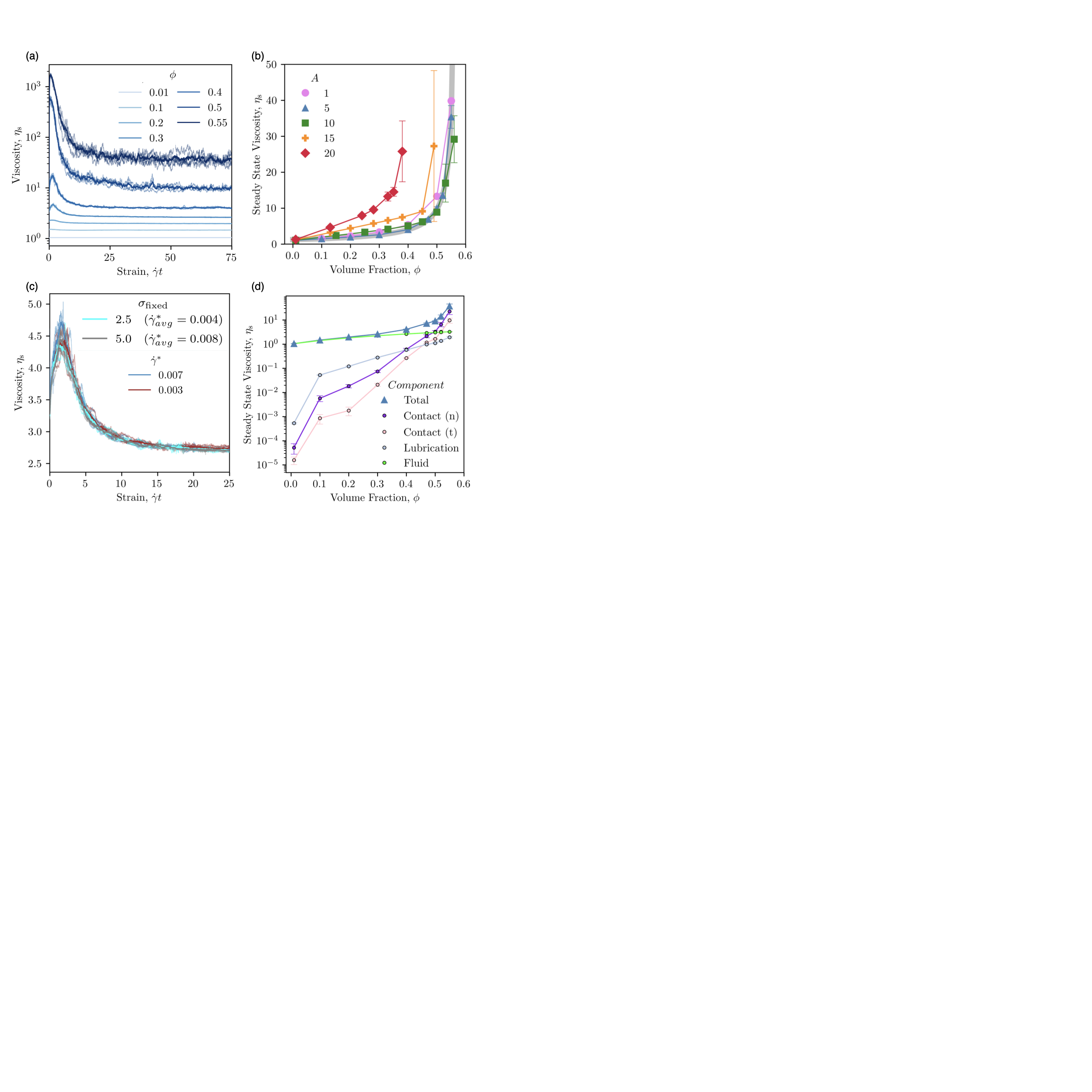}
        \caption{
        Simulation results showing the viscosity $\eta_\mathrm{s}$ of granular rod suspensions under simple shear flow.
        Shown are
        (a) typical transients of the suspension viscosity $\eta_\mathrm{s}$ plotted against accumulated strain $\dot{\gamma}t$ for aspect ratio $A=5$ at various volume fractions $\phi$, measured under a fixed shear rate of $\dot\gamma^* = 0.007$.
        Thinner lines show results for each individual realisation;
        (b) the dependence of the steady state suspension viscosity (obtained by  averaging the transient data over strains $\dot{\gamma}t>20$) on the volume fraction $\phi$ for aspect ratios $A=\{1,5,10,15, 20\}$.
        Data for $A=1$ (\emph{i.e.} spheres) are obtained from a separate model with equivalent physics~\cite{ness2023simulating,li2024simulating}.
        Shown in the solid gray line is the expression $\eta_\mathrm{s}=\iota(1-\phi/\phi_m)^{-\varsigma}$, with $\phi_m=0.57$, $\varsigma=0.97$ and $\iota=1.2$;
        (c) the transient suspension viscosity for $\phi=0.3$ and $A=5$, comparing fixed shear rate and fixed shear stress control modes to demonstrate their equivalence and the rate-independence of our model;
        (d) individual contributing components of the suspension viscosity for $A=5$,
        measured as a function of $\phi$. Included are the normal and tangential contact forces, the lubrication force, and the contributions from single-body hydrodynamics.
        All of the results presented represent averages taken over five realisations, each starting from distinct initial configurations.
        }
        \label{fig_4}
\end{figure}

We simulated aspect ratios $A=\{1,5, 10, 15,20\}$ to cover a broad range of particle shape, Figure~\ref{fig_3}.
Results for $A=1$ were obtained using a separate sphere-only model with equivalent physics~\cite{ness2023simulating,li2024simulating}.
For each aspect ratio we explored a range of volume fractions spanning from $\phi=0.01$
up to sufficiently high values to approach the jamming point,
which appears to decrease with $A$~\cite{tapia2017rheology}.
Transient and steady state viscosity data for each of these simulations are reported in Figure~\ref{fig_4}.

\begin{figure}
        \centering
        \includegraphics[trim = 1.0cm 53.6cm 50cm 2cm, clip,width=0.8\textwidth]{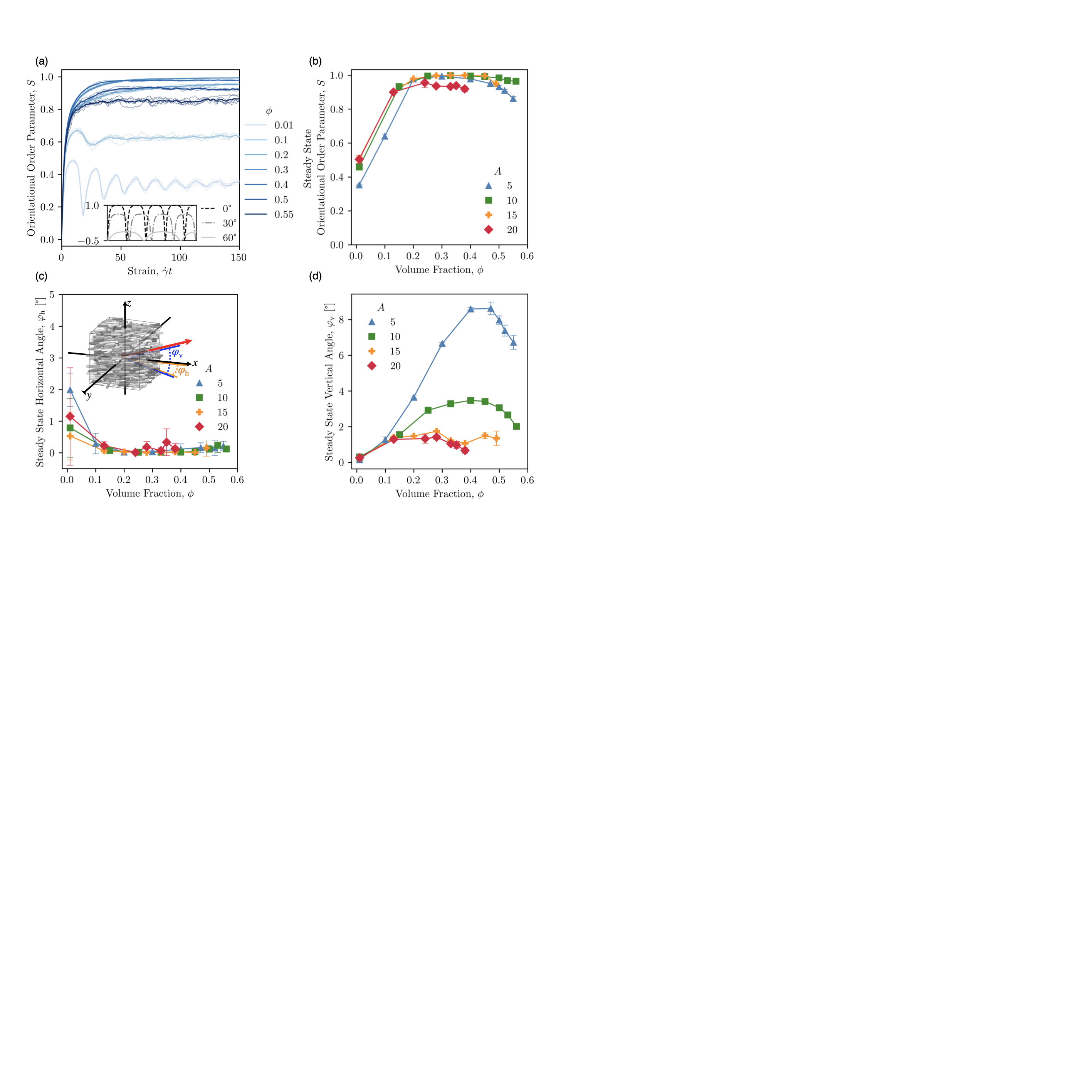}
        \caption{
        Simulation results showing the alignment of granular rods in suspension under simple shear flow. 
        Shown in (a) are transients of the orientational order parameter $S$ for $A=5$ and a range of volume fractions $\phi$ with the Inset showing the period of rotation of isolated rods ($\phi\to0$) under simple shear flow at angles to the flow-gradient ($xz$) plane = \{$0^\circ$, $30^\circ$, $60^\circ$\}. The steady states are plotted against volume fraction $\phi$ in (b), which additionally shows results for $A=\{10,15,20\}$.
        In (c) are the steady states of the angle $\varphi_\mathrm{h}$ between the projection onto the flow-vorticity ($xy$) plane of the principal direction of alignment obtained \emph{via} Equation~\ref{oop}, and the flow direction $x$. (d) The angle $\varphi_\mathrm{v}$ between the projection onto the flow-gradient ($xz$) plane of the principal direction of alignment and the flow direction $x$.
        The sketch in the Inset of (c) shows the definitions of $\varphi_\mathrm{h}$ and $\varphi_\mathrm{v}$.
        All of the results presented represent averages taken over five realisations, each starting from distinct initial configurations.
        }
        \label{fig_6}
\end{figure}

In the transients $\eta_\mathrm{s}(\dot{\gamma}t)$ for particles with $A=5$,
Figure~\ref{fig_4}(a),
there is an initial spike in the viscosity
occurring at the onset of shear around $\dot{\gamma}t=2$.
Comparing this transient to that of the alignment (Figure~\ref{fig_6}(a), discussed later),
it appears that the material has a short-lived higher viscosity associated with the initial random distribution of particle orientations.
This subsides once the particles start to align.
Thereafter the viscosity falls and remains fluctuating around a steady state value at large $\dot{\gamma}t$.
Notably,
the size of the initial spike in $\eta_\mathrm{s}$
relative to the steady state value is a growing function of $\phi$. 
This implies that the viscosity of the unsheared material may diverge more quickly with $\phi$ than the steady state viscosity,
suggesting microstructural dependence of the jamming point $\phi_m$. 
The strain scale needed to reach steady state also grows with $\phi$.
The viscosity at large strain increases systematically with increasing $\phi$,
consistent with a large body of experimental work on non-Brownian suspensions~\cite{guazzelli2018rheology}.

This trend of increasing viscosity with volume fraction is observed for all aspect ratios,
Figure~\ref{fig_4}(b).
In all cases the curves follow approximately the power law $\eta_\mathrm{s}\sim(1-\phi/\phi_m)^{-\varsigma}$ at large $\phi$,
with the example shown in Figure~\ref{fig_4}(b) for $A=5$ having $\phi_m=0.57$ and $\varsigma=0.97$.
Similar fits for the other curves (not plotted)
are poorer,
but suggest
$\phi_m \approx \{0.59,0.49,0.41\}$
for $A=\{10,15,20\}$.
The features of the viscosity,
namely the increase with respect to $\phi$ and the general variation of $\phi_m$ with $A$,
are qualitatively consistent with experiment~\cite{mueller2011effect,tapia2017rheology}.
At the microstructural level,
particles of higher aspect ratio can, for a given volume fraction, reach further into their neighbouring packing and can thus 
interact with more neighbours.
They can therefore reach the critical number of contacts required for mechanical stability at lower $\phi$,
explaining the relation between $\phi_m$ and $A$.
In our results we see a very similar jamming point for the three lowest aspect ratio particles $A=\{1,5,10\}$.
This is likely due to the very high degree of alignment (Figure~\ref{fig_6}(b),
discussed later) in these cases, meaning that the number of contacts is not being affected by the length (\emph{i.e.} two perfectly aligned $A=5$ rods would be roughly equivalent to one $A=10$ rod).
The higher aspect ratio rods experience much greater torques meaning they are constantly pushed out of this alignment and so their length continues to play a key role in their jamming point.
Additionally, higher aspect ratio particles contribute a greater stress to the fluid component (\emph{via} Equation~\ref{fluid_visc})
adding further to their increased viscosity at a given $\phi$.
We defer a comprehensive quantification and analysis of the jamming point and its relation to aspect ratio and particle alignment to a future study.

To verify the rate-independence of our model we ran simulations across both modes of operation, exploring a range of fixed shear rates and fixed shear stresses.
Representative results for $A=5$ and $\phi=0.3$,
Figure~\ref{fig_4}(c),
demonstrate that although the precise details of the transients differ,
the major features of the mechanical response are robust.
Importantly,
the steady state viscosity measured by averaging the transient over large strains
is independent of the shear rate.
Thus the viscosity divergences reported in Figure~\ref{fig_4}(b) are rate-independent,
consistent with the canonical framework of~\citet{boyer2011unifying} for granular suspensions.

Decomposing the viscosity into its contributions from each of the force terms detailed above,
we find a crossover in dominant microphysics occurring at intermediate $\phi$.
Again taking $A=5$ as a representative example,
we plot in Figure~\ref{fig_4}(d) the viscosity contributions from contact (normal and tangential) and lubrication forces as well as the fluid stress (Equation~\ref{fluid_visc}), as functions of $\phi$.
Here we see that the contact forces become negligible at small $\phi$ and instead the dominant contribution to the viscosity comes from the fluid component of the stress and secondarily from lubrication interactions.
As $\phi$ increases the contact components steadily rise and eventually dominate at $\phi \approx 0.5$.
The lubrication forces increase more slowly with $\phi$ and plateau beyond this point.
The value of the plateau is likely governed in part by our choice of $h_\mathrm{min}$,
which prevents the lubrication forces from diverging at particle contacts.
Above $\phi=0.5$ contacts dominate, with the normal component exceeding the tangential one, and their ratio likely governed by our choice of friction coefficient $\mu=0.5$.

\begin{figure}
        \centering
        \includegraphics[trim = 1.0cm 50.2cm 40cm 3cm, clip,width=0.9\textwidth]{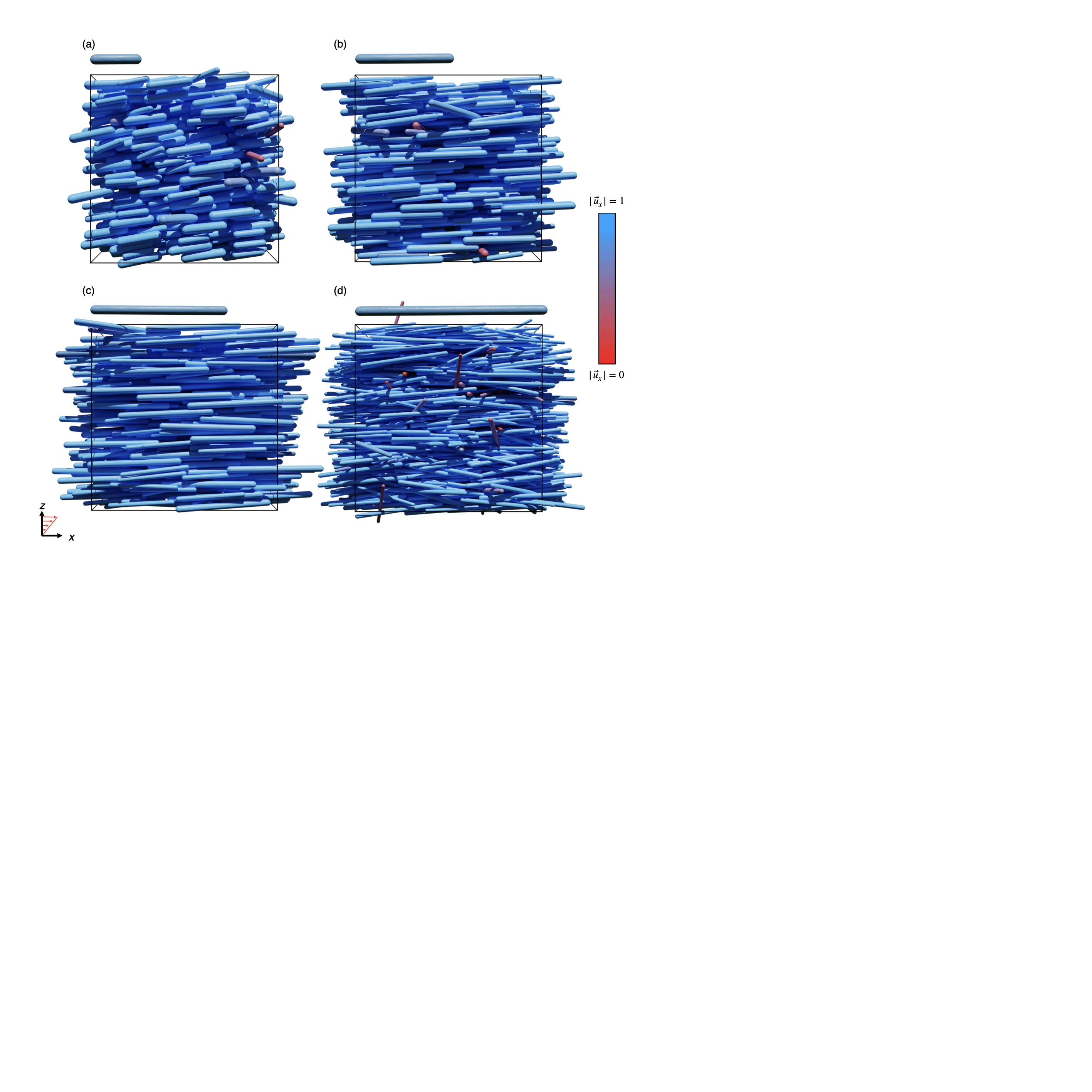}
        \caption{
Dense suspensions of granular rods under shear, showing example snapshots of the steady state alignment obtained at $\dot{\gamma}t=200$ for each aspect ratio.
Shown are the same cases as Figure~\ref{fig_3} with
        (a) $A=5, \phi =0.3$;
        (b) $A=10, \phi = 0.33$;
        (c) $A=15, \phi = 0.33$;
        (d) $A=20, \phi = 0.33$;
        red arrows and line sketched on the lower left coordinate system illustrate the directions of flow $x$ and velocity gradient $z$, while the colour bar (common to all panels) represents the $x$-component of the unit director $\vec{u}$ of each rod, with blue indicating perfect alignment along $x$.
        }
        \label{fig_5}
\end{figure}

\subsection{Effect of volume fraction and aspect ratio on particle alignment}

Shown in Figure~\ref{fig_5} are snapshots of simulations across $A$ illustrating the microstructural arrangements obtained after large strain.
In general we see that particles appear to strongly align along the flow direction.
We quantify these microstructural features using the order parameter $S$ defined above.
Additionally, we report two angles: $\varphi_{\mathrm{h}}$,
the angle formed between the flow direction $x$ and the projection onto the flow-vorticity ($xy$) plane of the eigenvector $\vec{u}_\mathrm{princ}$ associated with $S$;  and $\varphi_{\mathrm{v}}$, the angle formed between the flow direction $x$ and the projection onto the flow-gradient ($xz$) plane of $\vec{u}_\mathrm{princ}$.
Transients in system order $S$ for $A=5$ are shown in Figure~\ref{fig_6}(a) for a range of $\phi$.
In all cases there is, at small $\dot{\gamma}t$, an initial increase in $S$ from the random initial state with $S\approx0$,
implying a tendency toward alignment along the flow direction.
The strain over which this occurs maps onto that over which the viscosity was reported to initially spike in Figure~\ref{fig_4}.
This initial increase in order is followed by a slower evolution of $S$ with strain until steady states are reached.
% Remarkably, the transients in $S$ is often significantly longer than those in $\eta_\mathrm{s}$ for the same simulation.
% For instance,
% at $\phi=0.3$
% (green line in Figures~\ref{fig_3}(a),~\ref{fig_5}(a))
% the viscosity reaches a plateau at $\dot{\gamma}t\approx 5$,
% whereas $S_x$ continues to evolve until $\dot{\gamma}t\approx 100$.
Notably, at $\phi=0.01$ there is oscillatory behaviour in $S$ at large strain,
indicating that the packing is sparse enough to permit Jeffery-like orbits leading to periodically evolving alignment.
We verified that the period of the evolution in $S$ is indeed consistent with the expected Jeffery orbits by
separately simulating isolated rods in shear flow and measuring their alignment relative to the flow direction.
Doing so systematically for three inclinations out of the flow-gradient ($xz$) plane,
we obtain the transients plotted in Figure~\ref{fig_6}(a) Inset.
The periods of these isolated cases follow closely that of the sheared suspension at $\phi=0.01$,
albeit with extrema in the Inset varying across the full range of $S$ while those in the main panel of Figure~\ref{fig_6}(a) are more tightly bound around $S\approx0.35$.
Similar periodicity is initially evident at $\phi=0.1$, before dissipating at larger strains.

Computing the time averaged values of $S$ over $\dot{\gamma}t>200$, Figure~\ref{fig_6}(b), we see that there is a range of $\phi$ for which $S$ sharply increases, spanning approximately $\phi=0.01-0.2$ for $A=5$. At larger $\phi$ we see that $S$ maintains its very high order but does not increase any further. Further increasing $\phi$ then leads to a reduction in this steady state order. This indicates that the particles in the system become frustrated and are no longer as able to get into or maintain strong alignment with each other.
The same phenomenology in the ordering is observed for all other  aspect ratios. 

In addition to the extent of alignment,
it is important to characterise the direction of alignment.
The average alignment of the system in the $xy$ plane $\varphi_\mathrm{h}$ is shown in Figure~\ref{fig_6}(c).
Here an angle of $0^\circ$ corresponds to particles pointing along the direction of flow,
whereas an angle of $90^\circ$ would correspond to particles pointing into the vorticity direction.
In all cases the angle is $\approx 0$ with the $x$-axis,
indicating that the particles are preferentially aligning along the flow direction in a manner reported previously for sheared dry granular matter~\cite{borzsonyi2012orientational}. At very low volume fractions $\phi=0.01$ there is some deviation from zero, likely because these systems are not very well aligned ($S=0.3-0.5$) and do not have many contacts capable of adjusting a particle's trajectory, meaning that some particles will remain aligned along other directions executing kayaking-like orbits. This flow alignment is seen for all aspect ratios.

Shown in Figure~\ref{fig_6}(d) is the average inclination angle of the system $\varphi_\mathrm{v}$,
which takes values of $0^\circ$ for particles pointing along the flow direction and $90^\circ$ for particles pointing along the gradient direction.
Similar to the trend seen in $S$ we see a range of $\phi$ for which $\varphi_\mathrm{v}$ increases, spanning $\phi=0.01-0.4$, a larger range than $S$. This positive angle with the flow direction has again been previously reported for similar systems~\cite{borzsonyi2012orientational}. Past this point, we see that the inclination angle begins to decrease, corresponding with (and likely a consequence of) the decrease seen in $S$.  A similar trend is seen for the other aspect ratios. However as $A$ is increased, the maximum $\varphi_\mathrm{v}$ decreases and occurs at lower $\phi$. This appears as a flattening of the curve and suggests that at sufficiently high aspect ratio there will be negligible inclination.

\section{Conclusion and future outlook}

We have developed and implemented a numerical model containing a minimal set of physics describing dense suspensions of rod-shaped particles undergoing shear flow. The model incorporates standard forms for the contact,
drag and lift forces,
and modified lubrication forces that resolve a prior discontinuity.
It is accelerated by a novel dynamic scheme for computing the timestep.  
Predictions of the viscosity reveal a strong increase with the solid volume fraction,
which we varied across a broad range from $\phi=0.01$ to values approaching the jamming point $\phi_m$.
This jamming point,
at which the viscosity diverges, decreases with aspect ratio in the range $A=5$ to $20$,
consistent with experimental observation. 
Measurements of the shear-induced order show flow direction alignment increasing up to a  threshold volume fraction before slightly decreasing, a phenomenon seen for each aspect ratio.
Our model makes possible many future lines of study directly relevant to fluid mechanical problems in engineering applications,
for instance the effect of particle-particle friction coefficient on the viscosity and microstructure; the role of polydispersity both in the radius and aspect ratio of the particles;
the behaviour close to the jamming transition and its dependence on the particle alignment.
It also offers the opportunity to study other flow types such as pressure-imposed flow~\cite{etcheverry2023capillary} as well as non-linear velocity or stress profiles,
adapting ideas introduced recently for suspensions of spheres~\cite{saitoh2019nonlocal,bhowmik2024scaling,bhowmik2025unifying}.
More broadly it can be used to study problems at the intersection of jammed entangled granular matter~\cite{bhosale2022micromechanical} and dense suspension rheology~\cite{brown2011shear}.

\vspace{3mm}
\begin{acknowledgments}
A.D.~acknowledges financial support from Syngenta.
C.N.~acknowledges support from the Royal Academy of Engineering under the Research Fellowship scheme,
from the Leverhulme Trust under Research Project Grant RPG-2022-095
and from EPSRC Impact Acceleration Account EP/X525698/1. G.M. thanks NBIC/BBSRC/UKRI (grant no. BB/R012415/1) for financial support.
We thank Jason Butler, Eric Shaqfeh, John Brady, Holly Evans, Tyler Shendruk, Dan Hodgson, James Richards, Job Thijssen, Ben Goddard, and Diego Berzi for useful discussions.
The authors report no conflict of interest.
\end{acknowledgments}

\bibliography{apssamp}

\begin{thebibliography}{100}%
\makeatletter
\providecommand \@ifxundefined [1]{%
 \@ifx{#1\undefined}
}%
\providecommand \@ifnum [1]{%
 \ifnum #1\expandafter \@firstoftwo
 \else \expandafter \@secondoftwo
 \fi
}%
\providecommand \@ifx [1]{%
 \ifx #1\expandafter \@firstoftwo
 \else \expandafter \@secondoftwo
 \fi
}%
\providecommand \natexlab [1]{#1}%
\providecommand \enquote  [1]{``#1''}%
\providecommand \bibnamefont  [1]{#1}%
\providecommand \bibfnamefont [1]{#1}%
\providecommand \citenamefont [1]{#1}%
\providecommand \href@noop [0]{\@secondoftwo}%
\providecommand \href [0]{\begingroup \@sanitize@url \@href}%
\providecommand \@href[1]{\@@startlink{#1}\@@href}%
\providecommand \@@href[1]{\endgroup#1\@@endlink}%
\providecommand \@sanitize@url [0]{\catcode `\\12\catcode `\$12\catcode
  `\&12\catcode `\#12\catcode `\^12\catcode `\_12\catcode `\%12\relax}%
\providecommand \@@startlink[1]{}%
\providecommand \@@endlink[0]{}%
\providecommand \url  [0]{\begingroup\@sanitize@url \@url }%
\providecommand \@url [1]{\endgroup\@href {#1}{\urlprefix }}%
\providecommand \urlprefix  [0]{URL }%
\providecommand \Eprint [0]{\href }%
\providecommand \doibase [0]{https://doi.org/}%
\providecommand \selectlanguage [0]{\@gobble}%
\providecommand \bibinfo  [0]{\@secondoftwo}%
\providecommand \bibfield  [0]{\@secondoftwo}%
\providecommand \translation [1]{[#1]}%
\providecommand \BibitemOpen [0]{}%
\providecommand \bibitemStop [0]{}%
\providecommand \bibitemNoStop [0]{.\EOS\space}%
\providecommand \EOS [0]{\spacefactor3000\relax}%
\providecommand \BibitemShut  [1]{\csname bibitem#1\endcsname}%
\let\auto@bib@innerbib\@empty
%</preamble>
\bibitem [{\citenamefont {Butler}\ and\ \citenamefont
  {Snook}(2018)}]{butler2018microstructural}%
  \BibitemOpen
  \bibfield  {author} {\bibinfo {author} {\bibfnamefont {J.~E.}\ \bibnamefont
  {Butler}}\ and\ \bibinfo {author} {\bibfnamefont {B.}~\bibnamefont {Snook}},\
  }\bibfield  {title} {\bibinfo {title} {Microstructural dynamics and rheology
  of suspensions of rigid fibers},\ }\href@noop {} {\bibfield  {journal}
  {\bibinfo  {journal} {Annual Review of Fluid Mechanics}\ }\textbf {\bibinfo
  {volume} {50}},\ \bibinfo {pages} {299} (\bibinfo {year} {2018})}\BibitemShut
  {NoStop}%
\bibitem [{\citenamefont {Deshpande}\ and\ \citenamefont
  {Crosby}(2019)}]{deshpande2019logjams}%
  \BibitemOpen
  \bibfield  {author} {\bibinfo {author} {\bibfnamefont {N.~S.}\ \bibnamefont
  {Deshpande}}\ and\ \bibinfo {author} {\bibfnamefont {B.~T.}\ \bibnamefont
  {Crosby}},\ }\bibfield  {title} {\bibinfo {title} {Logjams are not jammed:
  measurements of log motions in {B}ig {C}reek, {I}daho},\ }\href@noop {}
  {\bibfield  {journal} {\bibinfo  {journal} {arXiv preprint arXiv:1911.01518}\
  } (\bibinfo {year} {2019})}\BibitemShut {NoStop}%
\bibitem [{\citenamefont {Cimarelli}\ \emph {et~al.}(2011)\citenamefont
  {Cimarelli}, \citenamefont {Costa}, \citenamefont {Mueller},\ and\
  \citenamefont {Mader}}]{cimarelli2011rheology}%
  \BibitemOpen
  \bibfield  {author} {\bibinfo {author} {\bibfnamefont {C.}~\bibnamefont
  {Cimarelli}}, \bibinfo {author} {\bibfnamefont {A.}~\bibnamefont {Costa}},
  \bibinfo {author} {\bibfnamefont {S.}~\bibnamefont {Mueller}},\ and\ \bibinfo
  {author} {\bibfnamefont {H.~M.}\ \bibnamefont {Mader}},\ }\bibfield  {title}
  {\bibinfo {title} {Rheology of magmas with bimodal crystal size and shape
  distributions: Insights from analog experiments},\ }\href@noop {} {\bibfield
  {journal} {\bibinfo  {journal} {Geochemistry, Geophysics, Geosystems}\
  }\textbf {\bibinfo {volume} {12}} (\bibinfo {year} {2011})}\BibitemShut
  {NoStop}%
\bibitem [{\citenamefont {Moitra}\ and\ \citenamefont
  {Gonnermann}(2015)}]{moitra2015effects}%
  \BibitemOpen
  \bibfield  {author} {\bibinfo {author} {\bibfnamefont {P.}~\bibnamefont
  {Moitra}}\ and\ \bibinfo {author} {\bibfnamefont {H.}~\bibnamefont
  {Gonnermann}},\ }\bibfield  {title} {\bibinfo {title} {Effects of crystal
  shape-and size-modality on magma rheology},\ }\href@noop {} {\bibfield
  {journal} {\bibinfo  {journal} {Geochemistry, Geophysics, Geosystems}\
  }\textbf {\bibinfo {volume} {16}},\ \bibinfo {pages} {1} (\bibinfo {year}
  {2015})}\BibitemShut {NoStop}%
\bibitem [{\citenamefont {Choo}\ \emph {et~al.}(2019)\citenamefont {Choo},
  \citenamefont {Mohd~Salleh}, \citenamefont {Kok},\ and\ \citenamefont
  {Matori}}]{choo2019review}%
  \BibitemOpen
  \bibfield  {author} {\bibinfo {author} {\bibfnamefont {T.~F.}\ \bibnamefont
  {Choo}}, \bibinfo {author} {\bibfnamefont {M.~A.}\ \bibnamefont
  {Mohd~Salleh}}, \bibinfo {author} {\bibfnamefont {K.~Y.}\ \bibnamefont
  {Kok}},\ and\ \bibinfo {author} {\bibfnamefont {K.~A.}\ \bibnamefont
  {Matori}},\ }\bibfield  {title} {\bibinfo {title} {A review on synthesis of
  mullite ceramics from industrial wastes},\ }\href@noop {} {\bibfield
  {journal} {\bibinfo  {journal} {Recycling}\ }\textbf {\bibinfo {volume}
  {4}},\ \bibinfo {pages} {39} (\bibinfo {year} {2019})}\BibitemShut {NoStop}%
\bibitem [{\citenamefont {Zhang}\ and\ \citenamefont
  {Darvell}(2010)}]{zhang2010synthesis}%
  \BibitemOpen
  \bibfield  {author} {\bibinfo {author} {\bibfnamefont {H.}~\bibnamefont
  {Zhang}}\ and\ \bibinfo {author} {\bibfnamefont {B.~W.}\ \bibnamefont
  {Darvell}},\ }\bibfield  {title} {\bibinfo {title} {Synthesis and
  characterization of hydroxyapatite whiskers by hydrothermal homogeneous
  precipitation using acetamide},\ }\href@noop {} {\bibfield  {journal}
  {\bibinfo  {journal} {Acta Biomaterialia}\ }\textbf {\bibinfo {volume} {6}},\
  \bibinfo {pages} {3216} (\bibinfo {year} {2010})}\BibitemShut {NoStop}%
\bibitem [{\citenamefont {Cao}\ \emph {et~al.}(2022)\citenamefont {Cao},
  \citenamefont {Liu}, \citenamefont {Li},\ and\ \citenamefont
  {Huang}}]{cao2022mechanical}%
  \BibitemOpen
  \bibfield  {author} {\bibinfo {author} {\bibfnamefont {K.}~\bibnamefont
  {Cao}}, \bibinfo {author} {\bibfnamefont {G.}~\bibnamefont {Liu}}, \bibinfo
  {author} {\bibfnamefont {H.}~\bibnamefont {Li}},\ and\ \bibinfo {author}
  {\bibfnamefont {Z.}~\bibnamefont {Huang}},\ }\bibfield  {title} {\bibinfo
  {title} {Mechanical properties and microstructure of calcium sulfate
  whisker-reinforced cement-based composites},\ }\href@noop {} {\bibfield
  {journal} {\bibinfo  {journal} {Materials}\ }\textbf {\bibinfo {volume}
  {15}},\ \bibinfo {pages} {947} (\bibinfo {year} {2022})}\BibitemShut
  {NoStop}%
\bibitem [{\citenamefont {Derakhshandeh}\ \emph {et~al.}(2011)\citenamefont
  {Derakhshandeh}, \citenamefont {Kerekes}, \citenamefont {Hatzikiriakos},\
  and\ \citenamefont {Bennington}}]{derakhshandeh2011rheology}%
  \BibitemOpen
  \bibfield  {author} {\bibinfo {author} {\bibfnamefont {B.}~\bibnamefont
  {Derakhshandeh}}, \bibinfo {author} {\bibfnamefont {R.~J.}\ \bibnamefont
  {Kerekes}}, \bibinfo {author} {\bibfnamefont {S.~G.}\ \bibnamefont
  {Hatzikiriakos}},\ and\ \bibinfo {author} {\bibfnamefont {C.~P.}\
  \bibnamefont {Bennington}},\ }\bibfield  {title} {\bibinfo {title} {Rheology
  of pulp fibre suspensions: A critical review},\ }\href@noop {} {\bibfield
  {journal} {\bibinfo  {journal} {Chemical Engineering Science}\ }\textbf
  {\bibinfo {volume} {66}},\ \bibinfo {pages} {3460} (\bibinfo {year}
  {2011})}\BibitemShut {NoStop}%
\bibitem [{\citenamefont {Jung}\ \emph {et~al.}(2025)\citenamefont {Jung},
  \citenamefont {Plumb-Reyes}, \citenamefont {Lin},\ and\ \citenamefont
  {Mahadevan}}]{jung2025entanglement}%
  \BibitemOpen
  \bibfield  {author} {\bibinfo {author} {\bibfnamefont {Y.}~\bibnamefont
  {Jung}}, \bibinfo {author} {\bibfnamefont {T.}~\bibnamefont {Plumb-Reyes}},
  \bibinfo {author} {\bibfnamefont {H.-Y.~G.}\ \bibnamefont {Lin}},\ and\
  \bibinfo {author} {\bibfnamefont {L.}~\bibnamefont {Mahadevan}},\ }\bibfield
  {title} {\bibinfo {title} {Entanglement transition in random rod packings},\
  }\href@noop {} {\bibfield  {journal} {\bibinfo  {journal} {Proceedings of the
  National Academy of Sciences}\ }\textbf {\bibinfo {volume} {122}},\ \bibinfo
  {pages} {e2401868122} (\bibinfo {year} {2025})}\BibitemShut {NoStop}%
\bibitem [{\citenamefont {Stokes}(1851)}]{stokes1851}%
  \BibitemOpen
  \bibfield  {author} {\bibinfo {author} {\bibfnamefont {G.~G.}\ \bibnamefont
  {Stokes}},\ }\bibfield  {title} {\bibinfo {title} {On the effect of the
  internal friction of fluids on the motion of pendulums},\ }\href@noop {}
  {\bibfield  {journal} {\bibinfo  {journal} {Transactions of the Cambridge
  Philosophical Society}\ }\textbf {\bibinfo {volume} {9}},\ \bibinfo {pages}
  {8–106} (\bibinfo {year} {1851})}\BibitemShut {NoStop}%
\bibitem [{\citenamefont {Einstein}(1911)}]{Einstein_1911}%
  \BibitemOpen
  \bibfield  {author} {\bibinfo {author} {\bibfnamefont {A.}~\bibnamefont
  {Einstein}},\ }\bibfield  {title} {\bibinfo {title} {Berichtigung zu meiner
  arbeit: Eine neue bestimmung der moleküldimensionen},\ }\href@noop {}
  {\bibfield  {journal} {\bibinfo  {journal} {Annalen der Physik}\ }\textbf
  {\bibinfo {volume} {34}},\ \bibinfo {pages} {591} (\bibinfo {year}
  {1911})}\BibitemShut {NoStop}%
\bibitem [{\citenamefont {Batchelor}\ and\ \citenamefont
  {Green}(1972)}]{batchelor1972determination}%
  \BibitemOpen
  \bibfield  {author} {\bibinfo {author} {\bibfnamefont {G.}~\bibnamefont
  {Batchelor}}\ and\ \bibinfo {author} {\bibfnamefont {J.}~\bibnamefont
  {Green}},\ }\bibfield  {title} {\bibinfo {title} {The determination of the
  bulk stress in a suspension of spherical particles to order c2},\ }\href@noop
  {} {\bibfield  {journal} {\bibinfo  {journal} {Journal of Fluid Mechanics}\
  }\textbf {\bibinfo {volume} {56}},\ \bibinfo {pages} {401} (\bibinfo {year}
  {1972})}\BibitemShut {NoStop}%
\bibitem [{\citenamefont {Krieger}\ and\ \citenamefont
  {Dougherty}(1959)}]{krieger1959mechanism}%
  \BibitemOpen
  \bibfield  {author} {\bibinfo {author} {\bibfnamefont {I.~M.}\ \bibnamefont
  {Krieger}}\ and\ \bibinfo {author} {\bibfnamefont {T.~J.}\ \bibnamefont
  {Dougherty}},\ }\bibfield  {title} {\bibinfo {title} {A mechanism for
  non-newtonian flow in suspensions of rigid spheres},\ }\href@noop {}
  {\bibfield  {journal} {\bibinfo  {journal} {Transactions of the Society of
  Rheology}\ }\textbf {\bibinfo {volume} {3}},\ \bibinfo {pages} {137}
  (\bibinfo {year} {1959})}\BibitemShut {NoStop}%
\bibitem [{\citenamefont {Boyer}\ \emph {et~al.}(2011)\citenamefont {Boyer},
  \citenamefont {Guazzelli},\ and\ \citenamefont
  {Pouliquen}}]{boyer2011unifying}%
  \BibitemOpen
  \bibfield  {author} {\bibinfo {author} {\bibfnamefont {F.}~\bibnamefont
  {Boyer}}, \bibinfo {author} {\bibfnamefont {{\'E}.}~\bibnamefont
  {Guazzelli}},\ and\ \bibinfo {author} {\bibfnamefont {O.}~\bibnamefont
  {Pouliquen}},\ }\bibfield  {title} {\bibinfo {title} {Unifying suspension and
  granular rheology},\ }\href@noop {} {\bibfield  {journal} {\bibinfo
  {journal} {Physical Review Letters}\ }\textbf {\bibinfo {volume} {107}},\
  \bibinfo {pages} {188301} (\bibinfo {year} {2011})}\BibitemShut {NoStop}%
\bibitem [{\citenamefont {Jeffery}(1922)}]{jeffery1922motion}%
  \BibitemOpen
  \bibfield  {author} {\bibinfo {author} {\bibfnamefont {G.~B.}\ \bibnamefont
  {Jeffery}},\ }\bibfield  {title} {\bibinfo {title} {The motion of ellipsoidal
  particles immersed in a viscous fluid},\ }\href@noop {} {\bibfield  {journal}
  {\bibinfo  {journal} {Proceedings of the Royal Society of London}\ }\textbf
  {\bibinfo {volume} {102}},\ \bibinfo {pages} {161} (\bibinfo {year}
  {1922})}\BibitemShut {NoStop}%
\bibitem [{\citenamefont {B{\"o}rzs{\"o}nyi}\ \emph
  {et~al.}(2012{\natexlab{a}})\citenamefont {B{\"o}rzs{\"o}nyi}, \citenamefont
  {Szab{\'o}}, \citenamefont {T{\"o}r{\"o}s}, \citenamefont {Wegner},
  \citenamefont {T{\"o}r{\"o}k}, \citenamefont {Somfai}, \citenamefont {Bien},\
  and\ \citenamefont {Stannarius}}]{borzsonyi2012orientational}%
  \BibitemOpen
  \bibfield  {author} {\bibinfo {author} {\bibfnamefont {T.}~\bibnamefont
  {B{\"o}rzs{\"o}nyi}}, \bibinfo {author} {\bibfnamefont {B.}~\bibnamefont
  {Szab{\'o}}}, \bibinfo {author} {\bibfnamefont {G.}~\bibnamefont
  {T{\"o}r{\"o}s}}, \bibinfo {author} {\bibfnamefont {S.}~\bibnamefont
  {Wegner}}, \bibinfo {author} {\bibfnamefont {J.}~\bibnamefont
  {T{\"o}r{\"o}k}}, \bibinfo {author} {\bibfnamefont {E.}~\bibnamefont
  {Somfai}}, \bibinfo {author} {\bibfnamefont {T.}~\bibnamefont {Bien}},\ and\
  \bibinfo {author} {\bibfnamefont {R.}~\bibnamefont {Stannarius}},\ }\bibfield
   {title} {\bibinfo {title} {Orientational order and alignment of elongated
  particles induced by shear},\ }\href@noop {} {\bibfield  {journal} {\bibinfo
  {journal} {Physical Review Letters}\ }\textbf {\bibinfo {volume} {108}},\
  \bibinfo {pages} {228302} (\bibinfo {year} {2012}{\natexlab{a}})}\BibitemShut
  {NoStop}%
\bibitem [{\citenamefont {B{\"o}rzs{\"o}nyi}\ \emph
  {et~al.}(2012{\natexlab{b}})\citenamefont {B{\"o}rzs{\"o}nyi}, \citenamefont
  {Szab{\'o}}, \citenamefont {Wegner}, \citenamefont {Harth}, \citenamefont
  {T{\"o}r{\"o}k}, \citenamefont {Somfai}, \citenamefont {Bien},\ and\
  \citenamefont {Stannarius}}]{borzsonyi2012shear}%
  \BibitemOpen
  \bibfield  {author} {\bibinfo {author} {\bibfnamefont {T.}~\bibnamefont
  {B{\"o}rzs{\"o}nyi}}, \bibinfo {author} {\bibfnamefont {B.}~\bibnamefont
  {Szab{\'o}}}, \bibinfo {author} {\bibfnamefont {S.}~\bibnamefont {Wegner}},
  \bibinfo {author} {\bibfnamefont {K.}~\bibnamefont {Harth}}, \bibinfo
  {author} {\bibfnamefont {J.}~\bibnamefont {T{\"o}r{\"o}k}}, \bibinfo {author}
  {\bibfnamefont {E.}~\bibnamefont {Somfai}}, \bibinfo {author} {\bibfnamefont
  {T.}~\bibnamefont {Bien}},\ and\ \bibinfo {author} {\bibfnamefont
  {R.}~\bibnamefont {Stannarius}},\ }\bibfield  {title} {\bibinfo {title}
  {Shear-induced alignment and dynamics of elongated granular particles},\
  }\href@noop {} {\bibfield  {journal} {\bibinfo  {journal} {Physical Review
  E}\ }\textbf {\bibinfo {volume} {86}},\ \bibinfo {pages} {051304} (\bibinfo
  {year} {2012}{\natexlab{b}})}\BibitemShut {NoStop}%
\bibitem [{\citenamefont {Franceschini}\ \emph {et~al.}(2011)\citenamefont
  {Franceschini}, \citenamefont {Filippidi}, \citenamefont {Guazzelli},\ and\
  \citenamefont {Pine}}]{franceschini2011transverse}%
  \BibitemOpen
  \bibfield  {author} {\bibinfo {author} {\bibfnamefont {A.}~\bibnamefont
  {Franceschini}}, \bibinfo {author} {\bibfnamefont {E.}~\bibnamefont
  {Filippidi}}, \bibinfo {author} {\bibfnamefont {E.}~\bibnamefont
  {Guazzelli}},\ and\ \bibinfo {author} {\bibfnamefont {D.~J.}\ \bibnamefont
  {Pine}},\ }\bibfield  {title} {\bibinfo {title} {Transverse alignment of
  fibers in a periodically sheared suspension: an absorbing phase transition
  with a slowly varying control parameter},\ }\href@noop {} {\bibfield
  {journal} {\bibinfo  {journal} {Physical Review Letters}\ }\textbf {\bibinfo
  {volume} {107}},\ \bibinfo {pages} {250603} (\bibinfo {year}
  {2011})}\BibitemShut {NoStop}%
\bibitem [{\citenamefont {Snook}\ \emph {et~al.}(2012)\citenamefont {Snook},
  \citenamefont {Guazzelli},\ and\ \citenamefont
  {Butler}}]{snook2012vorticity}%
  \BibitemOpen
  \bibfield  {author} {\bibinfo {author} {\bibfnamefont {B.}~\bibnamefont
  {Snook}}, \bibinfo {author} {\bibfnamefont {E.}~\bibnamefont {Guazzelli}},\
  and\ \bibinfo {author} {\bibfnamefont {J.~E.}\ \bibnamefont {Butler}},\
  }\bibfield  {title} {\bibinfo {title} {Vorticity alignment of rigid fibers in
  an oscillatory shear flow: Role of confinement},\ }\href@noop {} {\bibfield
  {journal} {\bibinfo  {journal} {Physics of Fluids}\ }\textbf {\bibinfo
  {volume} {24}} (\bibinfo {year} {2012})}\BibitemShut {NoStop}%
\bibitem [{\citenamefont {Franceschini}\ \emph {et~al.}(2014)\citenamefont
  {Franceschini}, \citenamefont {Filippidi}, \citenamefont {Guazzelli},\ and\
  \citenamefont {Pine}}]{franceschini2014dynamics}%
  \BibitemOpen
  \bibfield  {author} {\bibinfo {author} {\bibfnamefont {A.}~\bibnamefont
  {Franceschini}}, \bibinfo {author} {\bibfnamefont {E.}~\bibnamefont
  {Filippidi}}, \bibinfo {author} {\bibfnamefont {E.}~\bibnamefont
  {Guazzelli}},\ and\ \bibinfo {author} {\bibfnamefont {D.~J.}\ \bibnamefont
  {Pine}},\ }\bibfield  {title} {\bibinfo {title} {Dynamics of non-brownian
  fiber suspensions under periodic shear},\ }\href@noop {} {\bibfield
  {journal} {\bibinfo  {journal} {Soft Matter}\ }\textbf {\bibinfo {volume}
  {10}},\ \bibinfo {pages} {6722} (\bibinfo {year} {2014})}\BibitemShut
  {NoStop}%
\bibitem [{\citenamefont {Ralambotiana}\ \emph {et~al.}(1997)\citenamefont
  {Ralambotiana}, \citenamefont {Blanc},\ and\ \citenamefont
  {Chaouche}}]{ralambotiana1997viscosity}%
  \BibitemOpen
  \bibfield  {author} {\bibinfo {author} {\bibfnamefont {T.}~\bibnamefont
  {Ralambotiana}}, \bibinfo {author} {\bibfnamefont {R.}~\bibnamefont
  {Blanc}},\ and\ \bibinfo {author} {\bibfnamefont {M.}~\bibnamefont
  {Chaouche}},\ }\bibfield  {title} {\bibinfo {title} {Viscosity scaling in
  suspensions of non-brownian rodlike particles},\ }\href@noop {} {\bibfield
  {journal} {\bibinfo  {journal} {Physics of Fluids}\ }\textbf {\bibinfo
  {volume} {9}},\ \bibinfo {pages} {3588} (\bibinfo {year} {1997})}\BibitemShut
  {NoStop}%
\bibitem [{\citenamefont {Bibb{\'o}}(1987)}]{bibbo1987rheology}%
  \BibitemOpen
  \bibfield  {author} {\bibinfo {author} {\bibfnamefont {M.~A.}\ \bibnamefont
  {Bibb{\'o}}},\ }\emph {\bibinfo {title} {Rheology of semiconcentrated fiber
  suspensions}},\ \href@noop {} {\bibinfo {type} {Ph.{D}. thesis}},\ \bibinfo
  {school} {Massachusetts Institute of Technology} (\bibinfo {year}
  {1987})\BibitemShut {NoStop}%
\bibitem [{\citenamefont {Mackaplow}\ and\ \citenamefont
  {Shaqfeh}(1996)}]{mackaplow1996numerical}%
  \BibitemOpen
  \bibfield  {author} {\bibinfo {author} {\bibfnamefont {M.~B.}\ \bibnamefont
  {Mackaplow}}\ and\ \bibinfo {author} {\bibfnamefont {E.~S.}\ \bibnamefont
  {Shaqfeh}},\ }\bibfield  {title} {\bibinfo {title} {A numerical study of the
  rheological properties of suspensions of rigid, non-brownian fibres},\
  }\href@noop {} {\bibfield  {journal} {\bibinfo  {journal} {Journal of Fluid
  Mechanics}\ }\textbf {\bibinfo {volume} {329}},\ \bibinfo {pages} {155}
  (\bibinfo {year} {1996})}\BibitemShut {NoStop}%
\bibitem [{\citenamefont {Chaouche}\ and\ \citenamefont
  {Koch}(2001)}]{chaouche2001rheology}%
  \BibitemOpen
  \bibfield  {author} {\bibinfo {author} {\bibfnamefont {M.}~\bibnamefont
  {Chaouche}}\ and\ \bibinfo {author} {\bibfnamefont {D.~L.}\ \bibnamefont
  {Koch}},\ }\bibfield  {title} {\bibinfo {title} {Rheology of non-brownian
  rigid fiber suspensions with adhesive contacts},\ }\href@noop {} {\bibfield
  {journal} {\bibinfo  {journal} {Journal of Rheology}\ }\textbf {\bibinfo
  {volume} {45}},\ \bibinfo {pages} {369} (\bibinfo {year} {2001})}\BibitemShut
  {NoStop}%
\bibitem [{\citenamefont {Tapia}\ \emph {et~al.}(2017)\citenamefont {Tapia},
  \citenamefont {Shaikh}, \citenamefont {Butler}, \citenamefont {Pouliquen},\
  and\ \citenamefont {Guazzelli}}]{tapia2017rheology}%
  \BibitemOpen
  \bibfield  {author} {\bibinfo {author} {\bibfnamefont {F.}~\bibnamefont
  {Tapia}}, \bibinfo {author} {\bibfnamefont {S.}~\bibnamefont {Shaikh}},
  \bibinfo {author} {\bibfnamefont {J.~E.}\ \bibnamefont {Butler}}, \bibinfo
  {author} {\bibfnamefont {O.}~\bibnamefont {Pouliquen}},\ and\ \bibinfo
  {author} {\bibfnamefont {{\'E}.}~\bibnamefont {Guazzelli}},\ }\bibfield
  {title} {\bibinfo {title} {Rheology of concentrated suspensions of
  non-colloidal rigid fibres},\ }\href@noop {} {\bibfield  {journal} {\bibinfo
  {journal} {Journal of Fluid Mechanics}\ }\textbf {\bibinfo {volume} {827}},\
  \bibinfo {pages} {R5} (\bibinfo {year} {2017})}\BibitemShut {NoStop}%
\bibitem [{\citenamefont {Snook}\ \emph {et~al.}(2014)\citenamefont {Snook},
  \citenamefont {Davidson}, \citenamefont {Butler}, \citenamefont {Pouliquen},\
  and\ \citenamefont {Guazzelli}}]{snook2014normal}%
  \BibitemOpen
  \bibfield  {author} {\bibinfo {author} {\bibfnamefont {B.}~\bibnamefont
  {Snook}}, \bibinfo {author} {\bibfnamefont {L.~M.}\ \bibnamefont {Davidson}},
  \bibinfo {author} {\bibfnamefont {J.~E.}\ \bibnamefont {Butler}}, \bibinfo
  {author} {\bibfnamefont {O.}~\bibnamefont {Pouliquen}},\ and\ \bibinfo
  {author} {\bibfnamefont {E.}~\bibnamefont {Guazzelli}},\ }\bibfield  {title}
  {\bibinfo {title} {Normal stress differences in suspensions of rigid
  fibres},\ }\href@noop {} {\bibfield  {journal} {\bibinfo  {journal} {Journal
  of Fluid Mechanics}\ }\textbf {\bibinfo {volume} {758}},\ \bibinfo {pages}
  {486} (\bibinfo {year} {2014})}\BibitemShut {NoStop}%
\bibitem [{\citenamefont {Strand}\ \emph {et~al.}(1987)\citenamefont {Strand},
  \citenamefont {Kim},\ and\ \citenamefont {Karrila}}]{strand1987computation}%
  \BibitemOpen
  \bibfield  {author} {\bibinfo {author} {\bibfnamefont {S.~R.}\ \bibnamefont
  {Strand}}, \bibinfo {author} {\bibfnamefont {S.}~\bibnamefont {Kim}},\ and\
  \bibinfo {author} {\bibfnamefont {S.~J.}\ \bibnamefont {Karrila}},\
  }\bibfield  {title} {\bibinfo {title} {Computation of rheological properties
  of suspensions of rigid rods: stress growth after inception of steady shear
  flow},\ }\href@noop {} {\bibfield  {journal} {\bibinfo  {journal} {Journal of
  Non-Newtonian Fluid Mechanics}\ }\textbf {\bibinfo {volume} {24}},\ \bibinfo
  {pages} {311} (\bibinfo {year} {1987})}\BibitemShut {NoStop}%
\bibitem [{\citenamefont {Qui{\~n}ones}\ and\ \citenamefont
  {Olmsted}(2025)}]{quinones2025smoluchowski}%
  \BibitemOpen
  \bibfield  {author} {\bibinfo {author} {\bibfnamefont {C.}~\bibnamefont
  {Qui{\~n}ones}}\ and\ \bibinfo {author} {\bibfnamefont {P.~D.}\ \bibnamefont
  {Olmsted}},\ }\bibfield  {title} {\bibinfo {title} {A smoluchowski equation
  for a sheared suspension of frictionally interacting rods},\ }\href@noop {}
  {\bibfield  {journal} {\bibinfo  {journal} {arXiv preprint arXiv:2512.19149}\
  } (\bibinfo {year} {2025})}\BibitemShut {NoStop}%
\bibitem [{\citenamefont {Yamane}\ \emph {et~al.}(1994)\citenamefont {Yamane},
  \citenamefont {Kaneda},\ and\ \citenamefont {Doi}}]{yamane1994numerical}%
  \BibitemOpen
  \bibfield  {author} {\bibinfo {author} {\bibfnamefont {Y.}~\bibnamefont
  {Yamane}}, \bibinfo {author} {\bibfnamefont {Y.}~\bibnamefont {Kaneda}},\
  and\ \bibinfo {author} {\bibfnamefont {M.}~\bibnamefont {Doi}},\ }\bibfield
  {title} {\bibinfo {title} {Numerical simulation of semi-dilute suspensions of
  rodlike particles in shear flow},\ }\href@noop {} {\bibfield  {journal}
  {\bibinfo  {journal} {Journal of Non-Newtonian Fluid Mechanics}\ }\textbf
  {\bibinfo {volume} {54}},\ \bibinfo {pages} {405} (\bibinfo {year}
  {1994})}\BibitemShut {NoStop}%
\bibitem [{\citenamefont {Khan}\ \emph {et~al.}(2023)\citenamefont {Khan},
  \citenamefont {More}, \citenamefont {Banaei}, \citenamefont {Brandt},\ and\
  \citenamefont {Ardekani}}]{khan2023rheology}%
  \BibitemOpen
  \bibfield  {author} {\bibinfo {author} {\bibfnamefont {M.}~\bibnamefont
  {Khan}}, \bibinfo {author} {\bibfnamefont {R.~V.}\ \bibnamefont {More}},
  \bibinfo {author} {\bibfnamefont {A.~A.}\ \bibnamefont {Banaei}}, \bibinfo
  {author} {\bibfnamefont {L.}~\bibnamefont {Brandt}},\ and\ \bibinfo {author}
  {\bibfnamefont {A.~M.}\ \bibnamefont {Ardekani}},\ }\bibfield  {title}
  {\bibinfo {title} {Rheology of concentrated fiber suspensions with a
  load-dependent friction coefficient},\ }\href@noop {} {\bibfield  {journal}
  {\bibinfo  {journal} {Physical Review Fluids}\ }\textbf {\bibinfo {volume}
  {8}},\ \bibinfo {pages} {044301} (\bibinfo {year} {2023})}\BibitemShut
  {NoStop}%
\bibitem [{\citenamefont {Brady}\ and\ \citenamefont
  {Bossis}(1988)}]{brady1988stokesian}%
  \BibitemOpen
  \bibfield  {author} {\bibinfo {author} {\bibfnamefont {J.~F.}\ \bibnamefont
  {Brady}}\ and\ \bibinfo {author} {\bibfnamefont {G.}~\bibnamefont {Bossis}},\
  }\bibfield  {title} {\bibinfo {title} {Stokesian dynamics},\ }\href@noop {}
  {\bibfield  {journal} {\bibinfo  {journal} {Annual Review of Fluid
  Mechanics}\ }\textbf {\bibinfo {volume} {20}},\ \bibinfo {pages} {111}
  (\bibinfo {year} {1988})}\BibitemShut {NoStop}%
\bibitem [{\citenamefont {Phung}\ \emph {et~al.}(1996)\citenamefont {Phung},
  \citenamefont {Brady},\ and\ \citenamefont {Bossis}}]{phung1996stokesian}%
  \BibitemOpen
  \bibfield  {author} {\bibinfo {author} {\bibfnamefont {T.~N.}\ \bibnamefont
  {Phung}}, \bibinfo {author} {\bibfnamefont {J.~F.}\ \bibnamefont {Brady}},\
  and\ \bibinfo {author} {\bibfnamefont {G.}~\bibnamefont {Bossis}},\
  }\bibfield  {title} {\bibinfo {title} {Stokesian dynamics simulation of
  brownian suspensions},\ }\href@noop {} {\bibfield  {journal} {\bibinfo
  {journal} {Journal of Fluid Mechanics}\ }\textbf {\bibinfo {volume} {313}},\
  \bibinfo {pages} {181} (\bibinfo {year} {1996})}\BibitemShut {NoStop}%
\bibitem [{\citenamefont {Foss}\ and\ \citenamefont
  {Brady}(2000)}]{foss2000structure}%
  \BibitemOpen
  \bibfield  {author} {\bibinfo {author} {\bibfnamefont {D.~R.}\ \bibnamefont
  {Foss}}\ and\ \bibinfo {author} {\bibfnamefont {J.~F.}\ \bibnamefont
  {Brady}},\ }\bibfield  {title} {\bibinfo {title} {Structure, diffusion and
  rheology of brownian suspensions by stokesian dynamics simulation},\
  }\href@noop {} {\bibfield  {journal} {\bibinfo  {journal} {Journal of Fluid
  Mechanics}\ }\textbf {\bibinfo {volume} {407}},\ \bibinfo {pages} {167}
  (\bibinfo {year} {2000})}\BibitemShut {NoStop}%
\bibitem [{\citenamefont {Singh}\ and\ \citenamefont
  {Nott}(2000)}]{singh2000normal}%
  \BibitemOpen
  \bibfield  {author} {\bibinfo {author} {\bibfnamefont {A.}~\bibnamefont
  {Singh}}\ and\ \bibinfo {author} {\bibfnamefont {P.~R.}\ \bibnamefont
  {Nott}},\ }\bibfield  {title} {\bibinfo {title} {Normal stresses and
  microstructure in bounded sheared suspensions via stokesian dynamics
  simulations},\ }\href@noop {} {\bibfield  {journal} {\bibinfo  {journal}
  {Journal of Fluid Mechanics}\ }\textbf {\bibinfo {volume} {412}},\ \bibinfo
  {pages} {279} (\bibinfo {year} {2000})}\BibitemShut {NoStop}%
\bibitem [{\citenamefont {Kutteh}(2004)}]{kutteh2004methods}%
  \BibitemOpen
  \bibfield  {author} {\bibinfo {author} {\bibfnamefont {R.}~\bibnamefont
  {Kutteh}},\ }\bibfield  {title} {\bibinfo {title} {Methods for stokesian
  dynamics simulations of nonspherical particles and chains},\ }\href@noop {}
  {\bibfield  {journal} {\bibinfo  {journal} {Physical Review E}\ }\textbf
  {\bibinfo {volume} {69}},\ \bibinfo {pages} {011406} (\bibinfo {year}
  {2004})}\BibitemShut {NoStop}%
\bibitem [{\citenamefont {Mari}(2020)}]{mari2020shear}%
  \BibitemOpen
  \bibfield  {author} {\bibinfo {author} {\bibfnamefont {R.}~\bibnamefont
  {Mari}},\ }\bibfield  {title} {\bibinfo {title} {Shear thickening of
  suspensions of dimeric particles},\ }\href@noop {} {\bibfield  {journal}
  {\bibinfo  {journal} {Journal of Rheology}\ }\textbf {\bibinfo {volume}
  {64}},\ \bibinfo {pages} {239} (\bibinfo {year} {2020})}\BibitemShut
  {NoStop}%
\bibitem [{\citenamefont {Butler}\ and\ \citenamefont
  {Shaqfeh}(2002)}]{butler2002dynamic}%
  \BibitemOpen
  \bibfield  {author} {\bibinfo {author} {\bibfnamefont {J.~E.}\ \bibnamefont
  {Butler}}\ and\ \bibinfo {author} {\bibfnamefont {E.~S.}\ \bibnamefont
  {Shaqfeh}},\ }\bibfield  {title} {\bibinfo {title} {Dynamic simulations of
  the inhomogeneous sedimentation of rigid fibres},\ }\href@noop {} {\bibfield
  {journal} {\bibinfo  {journal} {Journal of Fluid Mechanics}\ }\textbf
  {\bibinfo {volume} {468}},\ \bibinfo {pages} {205} (\bibinfo {year}
  {2002})}\BibitemShut {NoStop}%
\bibitem [{\citenamefont {Fan}\ \emph {et~al.}(1998)\citenamefont {Fan},
  \citenamefont {Phan-Thien},\ and\ \citenamefont {Zheng}}]{fan1998direct}%
  \BibitemOpen
  \bibfield  {author} {\bibinfo {author} {\bibfnamefont {X.}~\bibnamefont
  {Fan}}, \bibinfo {author} {\bibfnamefont {N.}~\bibnamefont {Phan-Thien}},\
  and\ \bibinfo {author} {\bibfnamefont {R.}~\bibnamefont {Zheng}},\ }\bibfield
   {title} {\bibinfo {title} {A direct simulation of fibre suspensions},\
  }\href@noop {} {\bibfield  {journal} {\bibinfo  {journal} {Journal of
  Non-Newtonian Fluid Mechanics}\ }\textbf {\bibinfo {volume} {74}},\ \bibinfo
  {pages} {113} (\bibinfo {year} {1998})}\BibitemShut {NoStop}%
\bibitem [{\citenamefont {Durlofsky}\ \emph {et~al.}(1987)\citenamefont
  {Durlofsky}, \citenamefont {Brady},\ and\ \citenamefont
  {Bossis}}]{durlofsky1987dynamic}%
  \BibitemOpen
  \bibfield  {author} {\bibinfo {author} {\bibfnamefont {L.}~\bibnamefont
  {Durlofsky}}, \bibinfo {author} {\bibfnamefont {J.~F.}\ \bibnamefont
  {Brady}},\ and\ \bibinfo {author} {\bibfnamefont {G.}~\bibnamefont
  {Bossis}},\ }\bibfield  {title} {\bibinfo {title} {Dynamic simulation of
  hydrodynamically interacting particles},\ }\href@noop {} {\bibfield
  {journal} {\bibinfo  {journal} {Journal of Fluid Mechanics}\ }\textbf
  {\bibinfo {volume} {180}},\ \bibinfo {pages} {21} (\bibinfo {year}
  {1987})}\BibitemShut {NoStop}%
\bibitem [{\citenamefont {Cundall}\ and\ \citenamefont
  {Strack}(1979)}]{cundall1979discrete}%
  \BibitemOpen
  \bibfield  {author} {\bibinfo {author} {\bibfnamefont {P.~A.}\ \bibnamefont
  {Cundall}}\ and\ \bibinfo {author} {\bibfnamefont {O.~D.}\ \bibnamefont
  {Strack}},\ }\bibfield  {title} {\bibinfo {title} {A discrete numerical model
  for granular assemblies},\ }\href@noop {} {\bibfield  {journal} {\bibinfo
  {journal} {Geotechnique}\ }\textbf {\bibinfo {volume} {29}},\ \bibinfo
  {pages} {47} (\bibinfo {year} {1979})}\BibitemShut {NoStop}%
\bibitem [{\citenamefont {Plimpton}(1995)}]{plimpton1995lammps}%
  \BibitemOpen
  \bibfield  {author} {\bibinfo {author} {\bibfnamefont {S.}~\bibnamefont
  {Plimpton}},\ }\bibfield  {title} {\bibinfo {title} {Fast parallel algorithms
  for short-range molecular dynamics},\ }\href@noop {} {\bibfield  {journal}
  {\bibinfo  {journal} {Journal of Computational Physics}\ }\textbf {\bibinfo
  {volume} {117}},\ \bibinfo {pages} {1} (\bibinfo {year} {1995})}\BibitemShut
  {NoStop}%
\bibitem [{\citenamefont {Yousefian}\ and\ \citenamefont
  {Trulsson}(2022)}]{yousefian2022orientational}%
  \BibitemOpen
  \bibfield  {author} {\bibinfo {author} {\bibfnamefont {Z.}~\bibnamefont
  {Yousefian}}\ and\ \bibinfo {author} {\bibfnamefont {M.}~\bibnamefont
  {Trulsson}},\ }\bibfield  {title} {\bibinfo {title} {Orientational arrest in
  dense suspensions of elliptical particles under oscillatory shear flows},\
  }\href@noop {} {\bibfield  {journal} {\bibinfo  {journal} {Europhysics
  Letters}\ }\textbf {\bibinfo {volume} {136}},\ \bibinfo {pages} {36002}
  (\bibinfo {year} {2022})}\BibitemShut {NoStop}%
\bibitem [{\citenamefont {Zhang}\ and\ \citenamefont
  {Menon}(2025)}]{zhang2025hindered}%
  \BibitemOpen
  \bibfield  {author} {\bibinfo {author} {\bibfnamefont {Y.}~\bibnamefont
  {Zhang}}\ and\ \bibinfo {author} {\bibfnamefont {N.}~\bibnamefont {Menon}},\
  }\bibfield  {title} {\bibinfo {title} {Hindered stokesian settling of discs
  and rods},\ }\href@noop {} {\bibfield  {journal} {\bibinfo  {journal}
  {Physical Review Letters}\ }\textbf {\bibinfo {volume} {134}},\ \bibinfo
  {pages} {168202} (\bibinfo {year} {2025})}\BibitemShut {NoStop}%
\bibitem [{\citenamefont {Trulsson}(2021)}]{trulsson2021directional}%
  \BibitemOpen
  \bibfield  {author} {\bibinfo {author} {\bibfnamefont {M.}~\bibnamefont
  {Trulsson}},\ }\bibfield  {title} {\bibinfo {title} {Directional shear
  jamming of frictionless ellipses},\ }\href@noop {} {\bibfield  {journal}
  {\bibinfo  {journal} {Physical Review E}\ }\textbf {\bibinfo {volume}
  {104}},\ \bibinfo {pages} {044614} (\bibinfo {year} {2021})}\BibitemShut
  {NoStop}%
\bibitem [{\citenamefont {Bilotto}\ \emph {et~al.}(2025)\citenamefont
  {Bilotto}, \citenamefont {Trulsson},\ and\ \citenamefont
  {Molinari}}]{bilotto2025shear}%
  \BibitemOpen
  \bibfield  {author} {\bibinfo {author} {\bibfnamefont {J.}~\bibnamefont
  {Bilotto}}, \bibinfo {author} {\bibfnamefont {M.}~\bibnamefont {Trulsson}},\
  and\ \bibinfo {author} {\bibfnamefont {J.-F.}\ \bibnamefont {Molinari}},\
  }\bibfield  {title} {\bibinfo {title} {Shear flow of frictional spheroids:
  Comparison between elongated and flattened particles},\ }\href@noop {}
  {\bibfield  {journal} {\bibinfo  {journal} {Physical Review E}\ }\textbf
  {\bibinfo {volume} {112}},\ \bibinfo {pages} {045432} (\bibinfo {year}
  {2025})}\BibitemShut {NoStop}%
\bibitem [{\citenamefont {Ness}(2023)}]{ness2023simulating}%
  \BibitemOpen
  \bibfield  {author} {\bibinfo {author} {\bibfnamefont {C.}~\bibnamefont
  {Ness}},\ }\bibfield  {title} {\bibinfo {title} {Simulating dense,
  rate-independent suspension rheology using lammps},\ }\href@noop {}
  {\bibfield  {journal} {\bibinfo  {journal} {Computational Particle
  Mechanics}\ }\textbf {\bibinfo {volume} {10}},\ \bibinfo {pages} {2031}
  (\bibinfo {year} {2023})}\BibitemShut {NoStop}%
\bibitem [{\citenamefont {Dong}\ and\ \citenamefont
  {Trulsson}(2020)}]{dong2020unifying}%
  \BibitemOpen
  \bibfield  {author} {\bibinfo {author} {\bibfnamefont {J.}~\bibnamefont
  {Dong}}\ and\ \bibinfo {author} {\bibfnamefont {M.}~\bibnamefont
  {Trulsson}},\ }\bibfield  {title} {\bibinfo {title} {Unifying viscous and
  inertial regimes of discontinuous shear thickening suspensions},\ }\href@noop
  {} {\bibfield  {journal} {\bibinfo  {journal} {Journal of Rheology}\ }\textbf
  {\bibinfo {volume} {64}},\ \bibinfo {pages} {255} (\bibinfo {year}
  {2020})}\BibitemShut {NoStop}%
\bibitem [{\citenamefont {Ge}\ and\ \citenamefont
  {Brandt}(2020)}]{ge2020implementation}%
  \BibitemOpen
  \bibfield  {author} {\bibinfo {author} {\bibfnamefont {Z.}~\bibnamefont
  {Ge}}\ and\ \bibinfo {author} {\bibfnamefont {L.}~\bibnamefont {Brandt}},\
  }\bibfield  {title} {\bibinfo {title} {Implementation note on a minimal
  hybrid lubrication/granular dynamics model for dense suspensions},\
  }\href@noop {} {\bibfield  {journal} {\bibinfo  {journal} {arXiv preprint
  arXiv:2005.12755}\ } (\bibinfo {year} {2020})}\BibitemShut {NoStop}%
\bibitem [{\citenamefont {Gallier}\ \emph {et~al.}(2014)\citenamefont
  {Gallier}, \citenamefont {Lemaire}, \citenamefont {Peters},\ and\
  \citenamefont {Lobry}}]{gallier2014rheology}%
  \BibitemOpen
  \bibfield  {author} {\bibinfo {author} {\bibfnamefont {S.}~\bibnamefont
  {Gallier}}, \bibinfo {author} {\bibfnamefont {E.}~\bibnamefont {Lemaire}},
  \bibinfo {author} {\bibfnamefont {F.}~\bibnamefont {Peters}},\ and\ \bibinfo
  {author} {\bibfnamefont {L.}~\bibnamefont {Lobry}},\ }\bibfield  {title}
  {\bibinfo {title} {Rheology of sheared suspensions of rough frictional
  particles},\ }\href@noop {} {\bibfield  {journal} {\bibinfo  {journal}
  {Journal of Fluid Mechanics}\ }\textbf {\bibinfo {volume} {757}},\ \bibinfo
  {pages} {514} (\bibinfo {year} {2014})}\BibitemShut {NoStop}%
\bibitem [{\citenamefont {More}\ and\ \citenamefont
  {Ardekani}(2021)}]{more2021unifying}%
  \BibitemOpen
  \bibfield  {author} {\bibinfo {author} {\bibfnamefont {R.~V.}\ \bibnamefont
  {More}}\ and\ \bibinfo {author} {\bibfnamefont {A.~M.}\ \bibnamefont
  {Ardekani}},\ }\bibfield  {title} {\bibinfo {title} {Unifying disparate
  rate-dependent rheological regimes in non-brownian suspensions},\ }\href@noop
  {} {\bibfield  {journal} {\bibinfo  {journal} {Physical Review E}\ }\textbf
  {\bibinfo {volume} {103}},\ \bibinfo {pages} {062610} (\bibinfo {year}
  {2021})}\BibitemShut {NoStop}%
\bibitem [{\citenamefont {Guo}\ \emph {et~al.}(2012)\citenamefont {Guo},
  \citenamefont {Wassgren}, \citenamefont {Ketterhagen}, \citenamefont
  {Hancock}, \citenamefont {James},\ and\ \citenamefont
  {Curtis}}]{guo2012numerical}%
  \BibitemOpen
  \bibfield  {author} {\bibinfo {author} {\bibfnamefont {Y.}~\bibnamefont
  {Guo}}, \bibinfo {author} {\bibfnamefont {C.}~\bibnamefont {Wassgren}},
  \bibinfo {author} {\bibfnamefont {W.}~\bibnamefont {Ketterhagen}}, \bibinfo
  {author} {\bibfnamefont {B.}~\bibnamefont {Hancock}}, \bibinfo {author}
  {\bibfnamefont {B.}~\bibnamefont {James}},\ and\ \bibinfo {author}
  {\bibfnamefont {J.}~\bibnamefont {Curtis}},\ }\bibfield  {title} {\bibinfo
  {title} {A numerical study of granular shear flows of rod-like particles
  using the discrete element method},\ }\href@noop {} {\bibfield  {journal}
  {\bibinfo  {journal} {Journal of Fluid Mechanics}\ }\textbf {\bibinfo
  {volume} {713}},\ \bibinfo {pages} {1} (\bibinfo {year} {2012})}\BibitemShut
  {NoStop}%
\bibitem [{\citenamefont {Pol}\ \emph {et~al.}(2023)\citenamefont {Pol},
  \citenamefont {Artoni},\ and\ \citenamefont {Richard}}]{pol2023unified}%
  \BibitemOpen
  \bibfield  {author} {\bibinfo {author} {\bibfnamefont {A.}~\bibnamefont
  {Pol}}, \bibinfo {author} {\bibfnamefont {R.}~\bibnamefont {Artoni}},\ and\
  \bibinfo {author} {\bibfnamefont {P.}~\bibnamefont {Richard}},\ }\bibfield
  {title} {\bibinfo {title} {Unified scaling law for wall friction in laterally
  confined flows of shape anisotropic particles},\ }\href@noop {} {\bibfield
  {journal} {\bibinfo  {journal} {Physical Review Fluids}\ }\textbf {\bibinfo
  {volume} {8}},\ \bibinfo {pages} {084302} (\bibinfo {year}
  {2023})}\BibitemShut {NoStop}%
\bibitem [{\citenamefont {Nan}\ \emph {et~al.}(2015)\citenamefont {Nan},
  \citenamefont {Wang}, \citenamefont {Liu},\ and\ \citenamefont
  {Tang}}]{nan2015simulation}%
  \BibitemOpen
  \bibfield  {author} {\bibinfo {author} {\bibfnamefont {W.}~\bibnamefont
  {Nan}}, \bibinfo {author} {\bibfnamefont {Y.}~\bibnamefont {Wang}}, \bibinfo
  {author} {\bibfnamefont {Y.}~\bibnamefont {Liu}},\ and\ \bibinfo {author}
  {\bibfnamefont {H.}~\bibnamefont {Tang}},\ }\bibfield  {title} {\bibinfo
  {title} {Dem simulation of the packing of rodlike particles},\ }\href@noop {}
  {\bibfield  {journal} {\bibinfo  {journal} {Advanced Powder Technology}\
  }\textbf {\bibinfo {volume} {26}},\ \bibinfo {pages} {527} (\bibinfo {year}
  {2015})}\BibitemShut {NoStop}%
\bibitem [{\citenamefont {Guo}\ \emph {et~al.}(2015)\citenamefont {Guo},
  \citenamefont {Wassgren}, \citenamefont {Hancock}, \citenamefont
  {Ketterhagen},\ and\ \citenamefont {Curtis}}]{guo2015computational}%
  \BibitemOpen
  \bibfield  {author} {\bibinfo {author} {\bibfnamefont {Y.}~\bibnamefont
  {Guo}}, \bibinfo {author} {\bibfnamefont {C.}~\bibnamefont {Wassgren}},
  \bibinfo {author} {\bibfnamefont {B.}~\bibnamefont {Hancock}}, \bibinfo
  {author} {\bibfnamefont {W.}~\bibnamefont {Ketterhagen}},\ and\ \bibinfo
  {author} {\bibfnamefont {J.}~\bibnamefont {Curtis}},\ }\bibfield  {title}
  {\bibinfo {title} {Computational study of granular shear flows of dry
  flexible fibres using the discrete element method},\ }\href@noop {}
  {\bibfield  {journal} {\bibinfo  {journal} {Journal of Fluid Mechanics}\
  }\textbf {\bibinfo {volume} {775}},\ \bibinfo {pages} {24} (\bibinfo {year}
  {2015})}\BibitemShut {NoStop}%
\bibitem [{\citenamefont {Berzi}\ \emph {et~al.}(2016)\citenamefont {Berzi},
  \citenamefont {Thai-Quang}, \citenamefont {Guo},\ and\ \citenamefont
  {Curtis}}]{berzi2016stresses}%
  \BibitemOpen
  \bibfield  {author} {\bibinfo {author} {\bibfnamefont {D.}~\bibnamefont
  {Berzi}}, \bibinfo {author} {\bibfnamefont {N.}~\bibnamefont {Thai-Quang}},
  \bibinfo {author} {\bibfnamefont {Y.}~\bibnamefont {Guo}},\ and\ \bibinfo
  {author} {\bibfnamefont {J.}~\bibnamefont {Curtis}},\ }\bibfield  {title}
  {\bibinfo {title} {Stresses and orientational order in shearing flows of
  granular liquid crystals},\ }\href@noop {} {\bibfield  {journal} {\bibinfo
  {journal} {Physical Review E}\ }\textbf {\bibinfo {volume} {93}},\ \bibinfo
  {pages} {040901} (\bibinfo {year} {2016})}\BibitemShut {NoStop}%
\bibitem [{\citenamefont {Nath}\ and\ \citenamefont
  {Heussinger}(2019)}]{nath2019rheology}%
  \BibitemOpen
  \bibfield  {author} {\bibinfo {author} {\bibfnamefont {T.}~\bibnamefont
  {Nath}}\ and\ \bibinfo {author} {\bibfnamefont {C.}~\bibnamefont
  {Heussinger}},\ }\bibfield  {title} {\bibinfo {title} {Rheology in dense
  assemblies of spherocylinders: frictional vs. frictionless},\ }\href@noop {}
  {\bibfield  {journal} {\bibinfo  {journal} {The European Physical Journal E}\
  }\textbf {\bibinfo {volume} {42}},\ \bibinfo {pages} {1} (\bibinfo {year}
  {2019})}\BibitemShut {NoStop}%
\bibitem [{\citenamefont {Berzi}\ \emph {et~al.}(2022)\citenamefont {Berzi},
  \citenamefont {Buettner},\ and\ \citenamefont {Curtis}}]{berzi2022dense}%
  \BibitemOpen
  \bibfield  {author} {\bibinfo {author} {\bibfnamefont {D.}~\bibnamefont
  {Berzi}}, \bibinfo {author} {\bibfnamefont {K.~E.}\ \bibnamefont
  {Buettner}},\ and\ \bibinfo {author} {\bibfnamefont {J.~S.}\ \bibnamefont
  {Curtis}},\ }\bibfield  {title} {\bibinfo {title} {Dense shearing flows of
  soft, frictional cylinders},\ }\href@noop {} {\bibfield  {journal} {\bibinfo
  {journal} {Soft Matter}\ }\textbf {\bibinfo {volume} {18}},\ \bibinfo {pages}
  {80} (\bibinfo {year} {2022})}\BibitemShut {NoStop}%
\bibitem [{\citenamefont {Anzivino}\ \emph {et~al.}(2024)\citenamefont
  {Anzivino}, \citenamefont {Ness}, \citenamefont {Moussa},\ and\ \citenamefont
  {Zaccone}}]{anzivino2024shear}%
  \BibitemOpen
  \bibfield  {author} {\bibinfo {author} {\bibfnamefont {C.}~\bibnamefont
  {Anzivino}}, \bibinfo {author} {\bibfnamefont {C.}~\bibnamefont {Ness}},
  \bibinfo {author} {\bibfnamefont {A.~S.}\ \bibnamefont {Moussa}},\ and\
  \bibinfo {author} {\bibfnamefont {A.}~\bibnamefont {Zaccone}},\ }\bibfield
  {title} {\bibinfo {title} {Shear flow of non-brownian rod-sphere mixtures
  near jamming},\ }\href@noop {} {\bibfield  {journal} {\bibinfo  {journal}
  {Physical Review E}\ }\textbf {\bibinfo {volume} {109}},\ \bibinfo {pages}
  {L042601} (\bibinfo {year} {2024})}\BibitemShut {NoStop}%
\bibitem [{\citenamefont {Cheal}\ and\ \citenamefont
  {Ness}(2018)}]{cheal2018rheology}%
  \BibitemOpen
  \bibfield  {author} {\bibinfo {author} {\bibfnamefont {O.}~\bibnamefont
  {Cheal}}\ and\ \bibinfo {author} {\bibfnamefont {C.}~\bibnamefont {Ness}},\
  }\bibfield  {title} {\bibinfo {title} {Rheology of dense granular suspensions
  under extensional flow},\ }\href@noop {} {\bibfield  {journal} {\bibinfo
  {journal} {Journal of Rheology}\ }\textbf {\bibinfo {volume} {62}},\ \bibinfo
  {pages} {501} (\bibinfo {year} {2018})}\BibitemShut {NoStop}%
\bibitem [{\citenamefont {Seto}\ and\ \citenamefont
  {Giusteri}(2018)}]{seto2018normal}%
  \BibitemOpen
  \bibfield  {author} {\bibinfo {author} {\bibfnamefont {R.}~\bibnamefont
  {Seto}}\ and\ \bibinfo {author} {\bibfnamefont {G.~G.}\ \bibnamefont
  {Giusteri}},\ }\bibfield  {title} {\bibinfo {title} {Normal stress
  differences in dense suspensions},\ }\href@noop {} {\bibfield  {journal}
  {\bibinfo  {journal} {Journal of Fluid Mechanics}\ }\textbf {\bibinfo
  {volume} {857}},\ \bibinfo {pages} {200} (\bibinfo {year}
  {2018})}\BibitemShut {NoStop}%
\bibitem [{\citenamefont {Trulsson}\ \emph {et~al.}(2012)\citenamefont
  {Trulsson}, \citenamefont {Andreotti},\ and\ \citenamefont
  {Claudin}}]{trulsson2012transition}%
  \BibitemOpen
  \bibfield  {author} {\bibinfo {author} {\bibfnamefont {M.}~\bibnamefont
  {Trulsson}}, \bibinfo {author} {\bibfnamefont {B.}~\bibnamefont
  {Andreotti}},\ and\ \bibinfo {author} {\bibfnamefont {P.}~\bibnamefont
  {Claudin}},\ }\bibfield  {title} {\bibinfo {title} {Transition from the
  viscous to inertial regime in dense suspensions},\ }\href@noop {} {\bibfield
  {journal} {\bibinfo  {journal} {Physical Review Letters}\ }\textbf {\bibinfo
  {volume} {109}},\ \bibinfo {pages} {118305} (\bibinfo {year}
  {2012})}\BibitemShut {NoStop}%
\bibitem [{\citenamefont {Gan}\ and\ \citenamefont
  {Yu}(2020)}]{gan2020simulation}%
  \BibitemOpen
  \bibfield  {author} {\bibinfo {author} {\bibfnamefont {J.}~\bibnamefont
  {Gan}}\ and\ \bibinfo {author} {\bibfnamefont {A.}~\bibnamefont {Yu}},\
  }\bibfield  {title} {\bibinfo {title} {Dem simulation of the packing of
  cylindrical particles},\ }\href@noop {} {\bibfield  {journal} {\bibinfo
  {journal} {Granular Matter}\ }\textbf {\bibinfo {volume} {22}},\ \bibinfo
  {pages} {1} (\bibinfo {year} {2020})}\BibitemShut {NoStop}%
\bibitem [{\citenamefont {Lin}\ and\ \citenamefont
  {Han}(2002)}]{lin2002distance}%
  \BibitemOpen
  \bibfield  {author} {\bibinfo {author} {\bibfnamefont {A.}~\bibnamefont
  {Lin}}\ and\ \bibinfo {author} {\bibfnamefont {S.-P.}\ \bibnamefont {Han}},\
  }\bibfield  {title} {\bibinfo {title} {On the distance between two
  ellipsoids},\ }\href@noop {} {\bibfield  {journal} {\bibinfo  {journal} {SIAM
  Journal on Optimization}\ }\textbf {\bibinfo {volume} {13}},\ \bibinfo
  {pages} {298} (\bibinfo {year} {2002})}\BibitemShut {NoStop}%
\bibitem [{\citenamefont {Ch{\'e}ron}\ \emph {et~al.}(2024)\citenamefont
  {Ch{\'e}ron}, \citenamefont {Evrard},\ and\ \citenamefont {van
  Wachem}}]{cheron2024drag}%
  \BibitemOpen
  \bibfield  {author} {\bibinfo {author} {\bibfnamefont {V.}~\bibnamefont
  {Ch{\'e}ron}}, \bibinfo {author} {\bibfnamefont {F.}~\bibnamefont {Evrard}},\
  and\ \bibinfo {author} {\bibfnamefont {B.}~\bibnamefont {van Wachem}},\
  }\bibfield  {title} {\bibinfo {title} {Drag, lift and torque correlations for
  axi-symmetric rod-like non-spherical particles in locally linear shear
  flows},\ }\href@noop {} {\bibfield  {journal} {\bibinfo  {journal}
  {International Journal of Multiphase Flow}\ }\textbf {\bibinfo {volume}
  {171}},\ \bibinfo {pages} {104692} (\bibinfo {year} {2024})}\BibitemShut
  {NoStop}%
\bibitem [{\citenamefont {Feng}\ and\ \citenamefont
  {Michaelides}(2023)}]{feng2023general}%
  \BibitemOpen
  \bibfield  {author} {\bibinfo {author} {\bibfnamefont {Z.}~\bibnamefont
  {Feng}}\ and\ \bibinfo {author} {\bibfnamefont {E.~E.}\ \bibnamefont
  {Michaelides}},\ }\bibfield  {title} {\bibinfo {title} {A general and
  accurate correlation for the drag on spherocylinders},\ }\href@noop {}
  {\bibfield  {journal} {\bibinfo  {journal} {International Journal of
  Multiphase Flow}\ }\textbf {\bibinfo {volume} {168}},\ \bibinfo {pages}
  {104579} (\bibinfo {year} {2023})}\BibitemShut {NoStop}%
\bibitem [{\citenamefont {Happel}\ and\ \citenamefont
  {Brenner}(2012)}]{happel2012low}%
  \BibitemOpen
  \bibfield  {author} {\bibinfo {author} {\bibfnamefont {J.}~\bibnamefont
  {Happel}}\ and\ \bibinfo {author} {\bibfnamefont {H.}~\bibnamefont
  {Brenner}},\ }\href@noop {} {\emph {\bibinfo {title} {Low Reynolds number
  hydrodynamics: with special applications to particulate media}}},\
  Vol.~\bibinfo {volume} {1}\ (\bibinfo  {publisher} {Springer Science \&
  Business Media},\ \bibinfo {year} {2012})\BibitemShut {NoStop}%
\bibitem [{\citenamefont {Zhang}\ \emph {et~al.}(2001)\citenamefont {Zhang},
  \citenamefont {Ahmadi}, \citenamefont {Fan},\ and\ \citenamefont
  {McLaughlin}}]{zhang2001ellipsoidal}%
  \BibitemOpen
  \bibfield  {author} {\bibinfo {author} {\bibfnamefont {H.}~\bibnamefont
  {Zhang}}, \bibinfo {author} {\bibfnamefont {G.}~\bibnamefont {Ahmadi}},
  \bibinfo {author} {\bibfnamefont {F.-G.}\ \bibnamefont {Fan}},\ and\ \bibinfo
  {author} {\bibfnamefont {J.~B.}\ \bibnamefont {McLaughlin}},\ }\bibfield
  {title} {\bibinfo {title} {Ellipsoidal particles transport and deposition in
  turbulent channel flows},\ }\href@noop {} {\bibfield  {journal} {\bibinfo
  {journal} {International Journal of Multiphase Flow}\ }\textbf {\bibinfo
  {volume} {27}},\ \bibinfo {pages} {971} (\bibinfo {year} {2001})}\BibitemShut
  {NoStop}%
\bibitem [{\citenamefont {Brenner}(1963)}]{brenner1963stokes}%
  \BibitemOpen
  \bibfield  {author} {\bibinfo {author} {\bibfnamefont {H.}~\bibnamefont
  {Brenner}},\ }\bibfield  {title} {\bibinfo {title} {The stokes resistance of
  an arbitrary particle},\ }\href@noop {} {\bibfield  {journal} {\bibinfo
  {journal} {Chemical Engineering Science}\ }\textbf {\bibinfo {volume} {18}},\
  \bibinfo {pages} {1} (\bibinfo {year} {1963})}\BibitemShut {NoStop}%
\bibitem [{\citenamefont {Fan}\ and\ \citenamefont
  {Ahmadi}(1995)}]{fan1995sublayer}%
  \BibitemOpen
  \bibfield  {author} {\bibinfo {author} {\bibfnamefont {F.-G.}\ \bibnamefont
  {Fan}}\ and\ \bibinfo {author} {\bibfnamefont {G.}~\bibnamefont {Ahmadi}},\
  }\bibfield  {title} {\bibinfo {title} {A sublayer model for wall deposition
  of ellipsoidal particles in turbulent streams},\ }\href@noop {} {\bibfield
  {journal} {\bibinfo  {journal} {Journal of Aerosol Science}\ }\textbf
  {\bibinfo {volume} {26}},\ \bibinfo {pages} {813} (\bibinfo {year}
  {1995})}\BibitemShut {NoStop}%
\bibitem [{\citenamefont {Harper}\ and\ \citenamefont
  {Chang}(1968)}]{harper1968maximum}%
  \BibitemOpen
  \bibfield  {author} {\bibinfo {author} {\bibfnamefont {E.}~\bibnamefont
  {Harper}}\ and\ \bibinfo {author} {\bibfnamefont {I.-D.}\ \bibnamefont
  {Chang}},\ }\bibfield  {title} {\bibinfo {title} {Maximum dissipation
  resulting from lift in a slow viscous shear flow},\ }\href@noop {} {\bibfield
   {journal} {\bibinfo  {journal} {Journal of Fluid Mechanics}\ }\textbf
  {\bibinfo {volume} {33}},\ \bibinfo {pages} {209} (\bibinfo {year}
  {1968})}\BibitemShut {NoStop}%
\bibitem [{\citenamefont {Wang}\ \emph {et~al.}(2018)\citenamefont {Wang},
  \citenamefont {Yu}, \citenamefont {Luo}, \citenamefont {Xia},\ and\
  \citenamefont {Zong}}]{wang2018lift}%
  \BibitemOpen
  \bibfield  {author} {\bibinfo {author} {\bibfnamefont {J.}~\bibnamefont
  {Wang}}, \bibinfo {author} {\bibfnamefont {S.}~\bibnamefont {Yu}}, \bibinfo
  {author} {\bibfnamefont {S.}~\bibnamefont {Luo}}, \bibinfo {author}
  {\bibfnamefont {G.}~\bibnamefont {Xia}},\ and\ \bibinfo {author}
  {\bibfnamefont {L.}~\bibnamefont {Zong}},\ }\bibfield  {title} {\bibinfo
  {title} {Lift forces on axial symmetry particles rotating in a linear shear
  flow of a rarefied gas},\ }\href@noop {} {\bibfield  {journal} {\bibinfo
  {journal} {Physics of Fluids}\ }\textbf {\bibinfo {volume} {30}} (\bibinfo
  {year} {2018})}\BibitemShut {NoStop}%
\bibitem [{\citenamefont {Cui}\ \emph {et~al.}(2018)\citenamefont {Cui},
  \citenamefont {Decker}, \citenamefont {Janocha}, \citenamefont {K{\"o}rner},\
  and\ \citenamefont {Weigand}}]{cui2018constitutive}%
  \BibitemOpen
  \bibfield  {author} {\bibinfo {author} {\bibfnamefont {Y.}~\bibnamefont
  {Cui}}, \bibinfo {author} {\bibfnamefont {R.}~\bibnamefont {Decker}},
  \bibinfo {author} {\bibfnamefont {D.}~\bibnamefont {Janocha}}, \bibinfo
  {author} {\bibfnamefont {C.}~\bibnamefont {K{\"o}rner}},\ and\ \bibinfo
  {author} {\bibfnamefont {B.}~\bibnamefont {Weigand}},\ }\bibfield  {title}
  {\bibinfo {title} {On constitutive models for the momentum transfer to
  particles in fluid-dominated two-phase flows},\ }in\ \href@noop {} {\emph
  {\bibinfo {booktitle} {Advances in Mechanics of Materials and Structural
  Analysis: In Honor of Reinhold Kienzler}}}\ (\bibinfo  {publisher} {Springer
  International Publishing},\ \bibinfo {address} {Cham},\ \bibinfo {year}
  {2018})\ pp.\ \bibinfo {pages} {1--25}\BibitemShut {NoStop}%
\bibitem [{\citenamefont {Gallily}\ and\ \citenamefont
  {Cohen}(1979)}]{gallily1979orderly}%
  \BibitemOpen
  \bibfield  {author} {\bibinfo {author} {\bibfnamefont {I.}~\bibnamefont
  {Gallily}}\ and\ \bibinfo {author} {\bibfnamefont {A.-H.}\ \bibnamefont
  {Cohen}},\ }\bibfield  {title} {\bibinfo {title} {On the orderly nature of
  the motion of nonspherical aerosol particles. ii. inertial collision between
  a spherical large droplet and an axially symmetrical elongated particle},\
  }\href@noop {} {\bibfield  {journal} {\bibinfo  {journal} {Journal of Colloid
  and Interface Science}\ }\textbf {\bibinfo {volume} {68}},\ \bibinfo {pages}
  {338} (\bibinfo {year} {1979})}\BibitemShut {NoStop}%
\bibitem [{\citenamefont {Vega}\ and\ \citenamefont
  {Lago}(1994)}]{vega1994fast}%
  \BibitemOpen
  \bibfield  {author} {\bibinfo {author} {\bibfnamefont {C.}~\bibnamefont
  {Vega}}\ and\ \bibinfo {author} {\bibfnamefont {S.}~\bibnamefont {Lago}},\
  }\bibfield  {title} {\bibinfo {title} {A fast algorithm to evaluate the
  shortest distance between rods},\ }\href@noop {} {\bibfield  {journal}
  {\bibinfo  {journal} {Computers \& Chemistry}\ }\textbf {\bibinfo {volume}
  {18}},\ \bibinfo {pages} {55} (\bibinfo {year} {1994})}\BibitemShut {NoStop}%
\bibitem [{\citenamefont {Pournin}\ \emph {et~al.}(2005)\citenamefont
  {Pournin}, \citenamefont {Weber}, \citenamefont {Tsukahara}, \citenamefont
  {Ferrez}, \citenamefont {Ramaioli},\ and\ \citenamefont
  {Liebling}}]{pournin2005three}%
  \BibitemOpen
  \bibfield  {author} {\bibinfo {author} {\bibfnamefont {L.}~\bibnamefont
  {Pournin}}, \bibinfo {author} {\bibfnamefont {M.}~\bibnamefont {Weber}},
  \bibinfo {author} {\bibfnamefont {M.}~\bibnamefont {Tsukahara}}, \bibinfo
  {author} {\bibfnamefont {J.-A.}\ \bibnamefont {Ferrez}}, \bibinfo {author}
  {\bibfnamefont {M.}~\bibnamefont {Ramaioli}},\ and\ \bibinfo {author}
  {\bibfnamefont {T.~M.}\ \bibnamefont {Liebling}},\ }\bibfield  {title}
  {\bibinfo {title} {Three-dimensional distinct element simulation of
  spherocylinder crystallization},\ }\href@noop {} {\bibfield  {journal}
  {\bibinfo  {journal} {Granular Matter}\ }\textbf {\bibinfo {volume} {7}},\
  \bibinfo {pages} {119} (\bibinfo {year} {2005})}\BibitemShut {NoStop}%
\bibitem [{\citenamefont {Angklomkleaw}\ \emph {et~al.}(2019)\citenamefont
  {Angklomkleaw}, \citenamefont {Srinophakun},\ and\ \citenamefont
  {Tengpavadee}}]{angklomkleaw2019simulation}%
  \BibitemOpen
  \bibfield  {author} {\bibinfo {author} {\bibfnamefont {Y.}~\bibnamefont
  {Angklomkleaw}}, \bibinfo {author} {\bibfnamefont {T.~R.}\ \bibnamefont
  {Srinophakun}},\ and\ \bibinfo {author} {\bibfnamefont {S.}~\bibnamefont
  {Tengpavadee}},\ }\bibfield  {title} {\bibinfo {title} {Simulation of
  spherocylinder particle transport phenomena using discrete element method},\
  }\href@noop {} {\bibfield  {journal} {\bibinfo  {journal} {Asian Journal of
  Applied Sciences}\ }\textbf {\bibinfo {volume} {7}} (\bibinfo {year}
  {2019})}\BibitemShut {NoStop}%
\bibitem [{\citenamefont {Mahajan}\ \emph {et~al.}(2018)\citenamefont
  {Mahajan}, \citenamefont {Nijssen}, \citenamefont {Kuipers},\ and\
  \citenamefont {Padding}}]{mahajan2018non}%
  \BibitemOpen
  \bibfield  {author} {\bibinfo {author} {\bibfnamefont {V.~V.}\ \bibnamefont
  {Mahajan}}, \bibinfo {author} {\bibfnamefont {T.~M.}\ \bibnamefont
  {Nijssen}}, \bibinfo {author} {\bibfnamefont {J.}~\bibnamefont {Kuipers}},\
  and\ \bibinfo {author} {\bibfnamefont {J.~T.}\ \bibnamefont {Padding}},\
  }\bibfield  {title} {\bibinfo {title} {Non-spherical particles in a pseudo-2d
  fluidised bed: Modelling study},\ }\href@noop {} {\bibfield  {journal}
  {\bibinfo  {journal} {Chemical Engineering Science}\ }\textbf {\bibinfo
  {volume} {192}},\ \bibinfo {pages} {1105} (\bibinfo {year}
  {2018})}\BibitemShut {NoStop}%
\bibitem [{\citenamefont {Pournin}(2005)}]{pournin2005behavior}%
  \BibitemOpen
  \bibfield  {author} {\bibinfo {author} {\bibfnamefont {L.}~\bibnamefont
  {Pournin}},\ }\href@noop {} {\emph {\bibinfo {title} {On the behavior of
  spherical and non-spherical grain assemblies, its modeling and numerical
  simulation}}},\ \bibinfo {type} {Tech. Rep.}\ (\bibinfo  {institution}
  {EPFL},\ \bibinfo {year} {2005})\BibitemShut {NoStop}%
\bibitem [{\citenamefont {Constantin}(2015)}]{constantin2015aspect}%
  \BibitemOpen
  \bibfield  {author} {\bibinfo {author} {\bibfnamefont {D.}~\bibnamefont
  {Constantin}},\ }\bibfield  {title} {\bibinfo {title} {Why the aspect ratio?
  shape equivalence for the extinction spectra of gold nanoparticles},\
  }\href@noop {} {\bibfield  {journal} {\bibinfo  {journal} {The European
  Physical Journal E}\ }\textbf {\bibinfo {volume} {38}},\ \bibinfo {pages}
  {116} (\bibinfo {year} {2015})}\BibitemShut {NoStop}%
\bibitem [{\citenamefont {Janoschek}\ \emph {et~al.}(2013)\citenamefont
  {Janoschek}, \citenamefont {Harting},\ and\ \citenamefont
  {Toschi}}]{janoschek2013accurate}%
  \BibitemOpen
  \bibfield  {author} {\bibinfo {author} {\bibfnamefont {F.}~\bibnamefont
  {Janoschek}}, \bibinfo {author} {\bibfnamefont {J.}~\bibnamefont {Harting}},\
  and\ \bibinfo {author} {\bibfnamefont {F.}~\bibnamefont {Toschi}},\
  }\bibfield  {title} {\bibinfo {title} {Accurate lubrication corrections for
  spherical and non-spherical particles in discretized fluid simulations},\
  }\href@noop {} {\bibfield  {journal} {\bibinfo  {journal} {arXiv preprint
  arXiv:1308.6482}\ } (\bibinfo {year} {2013})}\BibitemShut {NoStop}%
\bibitem [{\citenamefont {Marschall}\ and\ \citenamefont
  {Teitel}(2019)}]{marschall2019shear}%
  \BibitemOpen
  \bibfield  {author} {\bibinfo {author} {\bibfnamefont {T.~A.}\ \bibnamefont
  {Marschall}}\ and\ \bibinfo {author} {\bibfnamefont {S.}~\bibnamefont
  {Teitel}},\ }\bibfield  {title} {\bibinfo {title} {Shear-driven flow of
  athermal, frictionless, spherocylinder suspensions in two dimensions: Stress,
  jamming, and contacts},\ }\href@noop {} {\bibfield  {journal} {\bibinfo
  {journal} {Physical Review E}\ }\textbf {\bibinfo {volume} {100}},\ \bibinfo
  {pages} {032906} (\bibinfo {year} {2019})}\BibitemShut {NoStop}%
\bibitem [{\citenamefont {Kuhn}(1933)}]{kuhn1933quantitative}%
  \BibitemOpen
  \bibfield  {author} {\bibinfo {author} {\bibfnamefont {W.}~\bibnamefont
  {Kuhn}},\ }\bibfield  {title} {\bibinfo {title} {{\"U}ber quantitative
  deutung der viskosit{\"a}t und str{\"o}mungsdoppelbrechung von
  suspensionen},\ }\href@noop {} {\bibfield  {journal} {\bibinfo  {journal}
  {Kolloid-Zeitschrift}\ }\textbf {\bibinfo {volume} {62}},\ \bibinfo {pages}
  {269} (\bibinfo {year} {1933})}\BibitemShut {NoStop}%
\bibitem [{\citenamefont {Dinh}\ and\ \citenamefont
  {Armstrong}(1984)}]{dinh1984rheological}%
  \BibitemOpen
  \bibfield  {author} {\bibinfo {author} {\bibfnamefont {S.~M.}\ \bibnamefont
  {Dinh}}\ and\ \bibinfo {author} {\bibfnamefont {R.~C.}\ \bibnamefont
  {Armstrong}},\ }\bibfield  {title} {\bibinfo {title} {A rheological equation
  of state for semiconcentrated fiber suspensions},\ }\href@noop {} {\bibfield
  {journal} {\bibinfo  {journal} {Journal of Rheology}\ }\textbf {\bibinfo
  {volume} {28}},\ \bibinfo {pages} {207} (\bibinfo {year} {1984})}\BibitemShut
  {NoStop}%
\bibitem [{\citenamefont {Lees}\ and\ \citenamefont
  {Edwards}(1972)}]{lees1972computer}%
  \BibitemOpen
  \bibfield  {author} {\bibinfo {author} {\bibfnamefont {A.~W.}\ \bibnamefont
  {Lees}}\ and\ \bibinfo {author} {\bibfnamefont {S.~F.}\ \bibnamefont
  {Edwards}},\ }\bibfield  {title} {\bibinfo {title} {The computer study of
  transport processes under extreme conditions},\ }\href@noop {} {\bibfield
  {journal} {\bibinfo  {journal} {Journal of Physics C: Solid State Physics}\
  }\textbf {\bibinfo {volume} {5}},\ \bibinfo {pages} {1921} (\bibinfo {year}
  {1972})}\BibitemShut {NoStop}%
\bibitem [{\citenamefont {VanderWerf}\ \emph {et~al.}(2018)\citenamefont
  {VanderWerf}, \citenamefont {Jin}, \citenamefont {Shattuck},\ and\
  \citenamefont {O’Hern}}]{VanderWerf_Jin_Shattuck_OHern_2018}%
  \BibitemOpen
  \bibfield  {author} {\bibinfo {author} {\bibfnamefont {K.}~\bibnamefont
  {VanderWerf}}, \bibinfo {author} {\bibfnamefont {W.}~\bibnamefont {Jin}},
  \bibinfo {author} {\bibfnamefont {M.~D.}\ \bibnamefont {Shattuck}},\ and\
  \bibinfo {author} {\bibfnamefont {C.~S.}\ \bibnamefont {O’Hern}},\
  }\bibfield  {title} {\bibinfo {title} {Hypostatic jammed packings of
  frictionless nonspherical particles},\ }\href@noop {} {\bibfield  {journal}
  {\bibinfo  {journal} {Physical Review E}\ }\textbf {\bibinfo {volume} {97}}
  (\bibinfo {year} {2018})}\BibitemShut {NoStop}%
\bibitem [{\citenamefont {O'hern}(2002)}]{Ohern2002-tw}%
  \BibitemOpen
  \bibfield  {author} {\bibinfo {author} {\bibfnamefont {C.~S.}\ \bibnamefont
  {O'hern}},\ }\bibfield  {title} {\bibinfo {title} {Random packings of
  frictionless particles},\ }\href@noop {} {\bibfield  {journal} {\bibinfo
  {journal} {Physical Review Letters}\ }\textbf {\bibinfo {volume} {88}}
  (\bibinfo {year} {2002})}\BibitemShut {NoStop}%
\bibitem [{\citenamefont {Zhang}(2009)}]{Zhang2009-yx}%
  \BibitemOpen
  \bibfield  {author} {\bibinfo {author} {\bibfnamefont {Z.}~\bibnamefont
  {Zhang}},\ }\bibfield  {title} {\bibinfo {title} {Thermal vestige of the
  zero-temperature jamming transition},\ }\href@noop {} {\bibfield  {journal}
  {\bibinfo  {journal} {Nature}\ }\textbf {\bibinfo {volume} {459}},\ \bibinfo
  {pages} {230} (\bibinfo {year} {2009})}\BibitemShut {NoStop}%
\bibitem [{\citenamefont {Singh}\ \emph {et~al.}(2024)\citenamefont {Singh},
  \citenamefont {Ness}, \citenamefont {Sharma}, \citenamefont {de~Pablo},\ and\
  \citenamefont {Jaeger}}]{singh2024rheology}%
  \BibitemOpen
  \bibfield  {author} {\bibinfo {author} {\bibfnamefont {A.}~\bibnamefont
  {Singh}}, \bibinfo {author} {\bibfnamefont {C.}~\bibnamefont {Ness}},
  \bibinfo {author} {\bibfnamefont {A.~K.}\ \bibnamefont {Sharma}}, \bibinfo
  {author} {\bibfnamefont {J.~J.}\ \bibnamefont {de~Pablo}},\ and\ \bibinfo
  {author} {\bibfnamefont {H.~M.}\ \bibnamefont {Jaeger}},\ }\bibfield  {title}
  {\bibinfo {title} {Rheology of bidisperse non-brownian suspensions},\
  }\href@noop {} {\bibfield  {journal} {\bibinfo  {journal} {Physical Review
  E}\ }\textbf {\bibinfo {volume} {110}},\ \bibinfo {pages} {034901} (\bibinfo
  {year} {2024})}\BibitemShut {NoStop}%
\bibitem [{\citenamefont {Mair}\ \emph {et~al.}(2002)\citenamefont {Mair},
  \citenamefont {Frye},\ and\ \citenamefont {Marone}}]{mair2002influence}%
  \BibitemOpen
  \bibfield  {author} {\bibinfo {author} {\bibfnamefont {K.}~\bibnamefont
  {Mair}}, \bibinfo {author} {\bibfnamefont {K.~M.}\ \bibnamefont {Frye}},\
  and\ \bibinfo {author} {\bibfnamefont {C.}~\bibnamefont {Marone}},\
  }\bibfield  {title} {\bibinfo {title} {Influence of grain characteristics on
  the friction of granular shear zones},\ }\href@noop {} {\bibfield  {journal}
  {\bibinfo  {journal} {Journal of Geophysical Research: Solid Earth}\ }\textbf
  {\bibinfo {volume} {107}},\ \bibinfo {pages} {ECV} (\bibinfo {year}
  {2002})}\BibitemShut {NoStop}%
\bibitem [{\citenamefont {Foerster}\ \emph {et~al.}(1994)\citenamefont
  {Foerster}, \citenamefont {Louge}, \citenamefont {Chang},\ and\ \citenamefont
  {Allia}}]{foerster1994measurements}%
  \BibitemOpen
  \bibfield  {author} {\bibinfo {author} {\bibfnamefont {S.~F.}\ \bibnamefont
  {Foerster}}, \bibinfo {author} {\bibfnamefont {M.~Y.}\ \bibnamefont {Louge}},
  \bibinfo {author} {\bibfnamefont {H.}~\bibnamefont {Chang}},\ and\ \bibinfo
  {author} {\bibfnamefont {K.}~\bibnamefont {Allia}},\ }\bibfield  {title}
  {\bibinfo {title} {Measurements of the collision properties of small
  spheres},\ }\href@noop {} {\bibfield  {journal} {\bibinfo  {journal} {Physics
  of Fluids}\ }\textbf {\bibinfo {volume} {6}},\ \bibinfo {pages} {1108}
  (\bibinfo {year} {1994})}\BibitemShut {NoStop}%
\bibitem [{\citenamefont {Joseph}\ \emph {et~al.}(2001)\citenamefont {Joseph},
  \citenamefont {Zenit}, \citenamefont {Hunt},\ and\ \citenamefont
  {Rosenwinkel}}]{joseph2001particle}%
  \BibitemOpen
  \bibfield  {author} {\bibinfo {author} {\bibfnamefont {G.}~\bibnamefont
  {Joseph}}, \bibinfo {author} {\bibfnamefont {R.}~\bibnamefont {Zenit}},
  \bibinfo {author} {\bibfnamefont {M.}~\bibnamefont {Hunt}},\ and\ \bibinfo
  {author} {\bibfnamefont {A.}~\bibnamefont {Rosenwinkel}},\ }\bibfield
  {title} {\bibinfo {title} {Particle--wall collisions in a viscous fluid},\
  }\href@noop {} {\bibfield  {journal} {\bibinfo  {journal} {Journal of Fluid
  Mechanics}\ }\textbf {\bibinfo {volume} {433}},\ \bibinfo {pages} {329}
  (\bibinfo {year} {2001})}\BibitemShut {NoStop}%
\bibitem [{\citenamefont {Li}\ \emph {et~al.}(2024)\citenamefont {Li},
  \citenamefont {Royer},\ and\ \citenamefont {Ness}}]{li2024simulating}%
  \BibitemOpen
  \bibfield  {author} {\bibinfo {author} {\bibfnamefont {X.}~\bibnamefont
  {Li}}, \bibinfo {author} {\bibfnamefont {J.~R.}\ \bibnamefont {Royer}},\ and\
  \bibinfo {author} {\bibfnamefont {C.}~\bibnamefont {Ness}},\ }\bibfield
  {title} {\bibinfo {title} {Simulating the rheology of dense suspensions using
  pairwise formulation of contact, lubrication and brownian forces},\
  }\href@noop {} {\bibfield  {journal} {\bibinfo  {journal} {Journal of Fluid
  Mechanics}\ }\textbf {\bibinfo {volume} {984}},\ \bibinfo {pages} {A67}
  (\bibinfo {year} {2024})}\BibitemShut {NoStop}%
\bibitem [{\citenamefont {Guazzelli}\ and\ \citenamefont
  {Pouliquen}(2018)}]{guazzelli2018rheology}%
  \BibitemOpen
  \bibfield  {author} {\bibinfo {author} {\bibfnamefont {{\'E}.}~\bibnamefont
  {Guazzelli}}\ and\ \bibinfo {author} {\bibfnamefont {O.}~\bibnamefont
  {Pouliquen}},\ }\bibfield  {title} {\bibinfo {title} {Rheology of dense
  granular suspensions},\ }\href@noop {} {\bibfield  {journal} {\bibinfo
  {journal} {Journal of Fluid Mechanics}\ }\textbf {\bibinfo {volume} {852}},\
  \bibinfo {pages} {P1} (\bibinfo {year} {2018})}\BibitemShut {NoStop}%
\bibitem [{\citenamefont {Mueller}\ \emph {et~al.}(2011)\citenamefont
  {Mueller}, \citenamefont {Llewellin},\ and\ \citenamefont
  {Mader}}]{mueller2011effect}%
  \BibitemOpen
  \bibfield  {author} {\bibinfo {author} {\bibfnamefont {S.}~\bibnamefont
  {Mueller}}, \bibinfo {author} {\bibfnamefont {E.}~\bibnamefont {Llewellin}},\
  and\ \bibinfo {author} {\bibfnamefont {H.}~\bibnamefont {Mader}},\ }\bibfield
   {title} {\bibinfo {title} {The effect of particle shape on suspension
  viscosity and implications for magmatic flows},\ }\href@noop {} {\bibfield
  {journal} {\bibinfo  {journal} {Geophysical Research Letters}\ }\textbf
  {\bibinfo {volume} {38}} (\bibinfo {year} {2011})}\BibitemShut {NoStop}%
\bibitem [{\citenamefont {Etcheverry}\ \emph {et~al.}(2023)\citenamefont
  {Etcheverry}, \citenamefont {Forterre},\ and\ \citenamefont
  {Metzger}}]{etcheverry2023capillary}%
  \BibitemOpen
  \bibfield  {author} {\bibinfo {author} {\bibfnamefont {B.}~\bibnamefont
  {Etcheverry}}, \bibinfo {author} {\bibfnamefont {Y.}~\bibnamefont
  {Forterre}},\ and\ \bibinfo {author} {\bibfnamefont {B.}~\bibnamefont
  {Metzger}},\ }\bibfield  {title} {\bibinfo {title} {Capillary-stress
  controlled rheometer reveals the dual rheology of shear-thickening
  suspensions},\ }\href@noop {} {\bibfield  {journal} {\bibinfo  {journal}
  {Physical Review X}\ }\textbf {\bibinfo {volume} {13}},\ \bibinfo {pages}
  {011024} (\bibinfo {year} {2023})}\BibitemShut {NoStop}%
\bibitem [{\citenamefont {Saitoh}\ and\ \citenamefont
  {Tighe}(2019)}]{saitoh2019nonlocal}%
  \BibitemOpen
  \bibfield  {author} {\bibinfo {author} {\bibfnamefont {K.}~\bibnamefont
  {Saitoh}}\ and\ \bibinfo {author} {\bibfnamefont {B.~P.}\ \bibnamefont
  {Tighe}},\ }\bibfield  {title} {\bibinfo {title} {Nonlocal effects in
  inhomogeneous flows of soft athermal disks},\ }\href@noop {} {\bibfield
  {journal} {\bibinfo  {journal} {Physical Review Letters}\ }\textbf {\bibinfo
  {volume} {122}},\ \bibinfo {pages} {188001} (\bibinfo {year}
  {2019})}\BibitemShut {NoStop}%
\bibitem [{\citenamefont {Bhowmik}\ and\ \citenamefont
  {Ness}(2024)}]{bhowmik2024scaling}%
  \BibitemOpen
  \bibfield  {author} {\bibinfo {author} {\bibfnamefont {B.~P.}\ \bibnamefont
  {Bhowmik}}\ and\ \bibinfo {author} {\bibfnamefont {C.}~\bibnamefont {Ness}},\
  }\bibfield  {title} {\bibinfo {title} {Scaling description of frictionless
  dense suspensions under inhomogeneous flow},\ }\href@noop {} {\bibfield
  {journal} {\bibinfo  {journal} {Physical Review Letters}\ }\textbf {\bibinfo
  {volume} {132}},\ \bibinfo {pages} {118203} (\bibinfo {year}
  {2024})}\BibitemShut {NoStop}%
\bibitem [{\citenamefont {Bhowmik}\ and\ \citenamefont
  {Ness}(2025)}]{bhowmik2025unifying}%
  \BibitemOpen
  \bibfield  {author} {\bibinfo {author} {\bibfnamefont {B.~P.}\ \bibnamefont
  {Bhowmik}}\ and\ \bibinfo {author} {\bibfnamefont {C.}~\bibnamefont {Ness}},\
  }\bibfield  {title} {\bibinfo {title} {Unifying homogeneous and inhomogeneous
  rheology of dense suspensions},\ }\href@noop {} {\bibfield  {journal}
  {\bibinfo  {journal} {Journal of Rheology}\ }\textbf {\bibinfo {volume}
  {69}},\ \bibinfo {pages} {423} (\bibinfo {year} {2025})}\BibitemShut
  {NoStop}%
\bibitem [{\citenamefont {Bhosale}\ \emph {et~al.}(2022)\citenamefont
  {Bhosale}, \citenamefont {Weiner}, \citenamefont {Butler}, \citenamefont
  {Kim}, \citenamefont {Gazzola},\ and\ \citenamefont
  {King}}]{bhosale2022micromechanical}%
  \BibitemOpen
  \bibfield  {author} {\bibinfo {author} {\bibfnamefont {Y.}~\bibnamefont
  {Bhosale}}, \bibinfo {author} {\bibfnamefont {N.}~\bibnamefont {Weiner}},
  \bibinfo {author} {\bibfnamefont {A.}~\bibnamefont {Butler}}, \bibinfo
  {author} {\bibfnamefont {S.~H.}\ \bibnamefont {Kim}}, \bibinfo {author}
  {\bibfnamefont {M.}~\bibnamefont {Gazzola}},\ and\ \bibinfo {author}
  {\bibfnamefont {H.}~\bibnamefont {King}},\ }\bibfield  {title} {\bibinfo
  {title} {Micromechanical origin of plasticity and hysteresis in nestlike
  packings},\ }\href@noop {} {\bibfield  {journal} {\bibinfo  {journal}
  {Physical Review Letters}\ }\textbf {\bibinfo {volume} {128}},\ \bibinfo
  {pages} {198003} (\bibinfo {year} {2022})}\BibitemShut {NoStop}%
\bibitem [{\citenamefont {Brown}\ \emph {et~al.}(2011)\citenamefont {Brown},
  \citenamefont {Zhang}, \citenamefont {Forman}, \citenamefont {Maynor},
  \citenamefont {Betts}, \citenamefont {DeSimone},\ and\ \citenamefont
  {Jaeger}}]{brown2011shear}%
  \BibitemOpen
  \bibfield  {author} {\bibinfo {author} {\bibfnamefont {E.}~\bibnamefont
  {Brown}}, \bibinfo {author} {\bibfnamefont {H.}~\bibnamefont {Zhang}},
  \bibinfo {author} {\bibfnamefont {N.~A.}\ \bibnamefont {Forman}}, \bibinfo
  {author} {\bibfnamefont {B.~W.}\ \bibnamefont {Maynor}}, \bibinfo {author}
  {\bibfnamefont {D.~E.}\ \bibnamefont {Betts}}, \bibinfo {author}
  {\bibfnamefont {J.~M.}\ \bibnamefont {DeSimone}},\ and\ \bibinfo {author}
  {\bibfnamefont {H.~M.}\ \bibnamefont {Jaeger}},\ }\bibfield  {title}
  {\bibinfo {title} {Shear thickening and jamming in densely packed suspensions
  of different particle shapes},\ }\href@noop {} {\bibfield  {journal}
  {\bibinfo  {journal} {Physical Review E}\ }\textbf {\bibinfo {volume} {84}},\
  \bibinfo {pages} {031408} (\bibinfo {year} {2011})}\BibitemShut {NoStop}%
\end{thebibliography}%
\end{document}